\documentclass[aps,prd,preprint,superscriptaddress,showpacs]{revtex4}
\usepackage{slashed}
\usepackage{graphicx}
\usepackage{verbatim}
\usepackage{ulem}
\usepackage{color}
\usepackage{amssymb}
\definecolor{My_red}        {cmyk}{0.00,1.00,1.00,0.20}

\newcommand{\bmat}{\left(\begin{array}}
\newcommand{\emat}{\end{array}\right)}
\newcommand{\beq}{\begin{equation}}
\newcommand{\eeq}{\end{equation}}
\newcommand{\wt}{\widetilde}

\def\ra{\rightarrow}

\def\ld{\lambda}
\def\f{\frac}

\def\bwt{\begin{widetext}}
\def\ewt{\end{widetext}}
\def\be{\begin{equation}}
\def\ee{\end{equation}}
\def\bea{\begin{eqnarray}}
\def\eea{\end{eqnarray}}
\def\bean{\begin{eqnarray*}}
\def\eean{\end{eqnarray*}}
\def\bary{\begin{array}}
\def\eary{\end{array}}
\def\bit{\begin{itemize}}
\def\eit{\end{itemize}}

\def\ra{\rightarrow}

\def\ld{\lambda}

\def\su5u1{SU(5) \times U(1)}
\def\fsu5u1{SU(5) \times U(1)'}
\def\so10{SO(10)}
\def\sq20{SO(10) \times SO(10)}

\def\ra{\rightarrow}

\def\ld{\lambda}
\def\f{\frac}

\def\L{\left(}
\def\R{\right)}

\def\ra{\rightarrow}

\def\ld{\lambda}

\def\su5u1{SU(5) \times U(1)}
\def\fsu5u1{SU(5) \times U(1)'}
\def\so10{SO(10)}
\def\sq20{SO(10) \times SO(10)}

\usepackage[centertags]{amsmath}
\usepackage{amssymb}

\begin{document}

\title{Implications of Higgs Sterility for the Higgs and Stop Sectors}

\author{Jun Guo}
\email{hustgj@itp.ac.cn}
 \affiliation{State Key Laboratory of
Theoretical Physics and Kavli Institute for Theoretical Physics
China (KITPC), Institute of Theoretical Physics, Chinese Academy of
Sciences, Beijing 100190, P. R. China}

\author{Zhaofeng Kang}
\email[E-mail: ]{zhaofengkang@gmail.com}
\affiliation{Center for High-Energy Physics, Peking University, Beijing, 100871, P. R. China}

\author{Jinmin Li}
\email{jmli@itp.ac.cn}

\affiliation{State Key Laboratory of Theoretical Physics
and Kavli Institute for Theoretical Physics China (KITPC),
Institute of Theoretical Physics, Chinese Academy of Sciences,
Beijing 100190, P. R. China}

\author{Tianjun Li}
\email{tli@itp.ac.cn}

\affiliation{State Key Laboratory of Theoretical Physics
and Kavli Institute for Theoretical Physics China (KITPC),
Institute of Theoretical Physics, Chinese Academy of Sciences,
Beijing 100190, P. R. China}

\affiliation{School of Physical Electronics,
University of Electronic Science and Technology of China,
Chengdu 610054, P. R. China}

\begin{abstract}

The LHC data implies that the newly discovered Higgs boson $h$ may
be sterile (highly SM-like). In supersymmetric SMs (SSMs), Higgs
couplings are often modified by Higgs mixing and stop loop
corrections, so we study the Higgs sterility in the Higgs and stop
sectors in two SSMs: (I) The Minimal SSM (MSSM). In the nearly
decoupling region, the doublet-doublet mixing effect can only
enhance $C_{hb\bar b}$ by $2m_Z^2/M_A^2$. Sterility places $M_A
\gtrsim 900$ GeV. But it hardly constrains the stop sector due to
the heaviness of Higgs boson mass $m_h$; (II) The next to MSSM
(NMSSM). In the presence of doublet-singlet mixing, the mixing
structure is complicated. We find a simple approximation to
understand Higgs sterility and its implications, says the amount of
pushing-up $m_h\lesssim$ 5 GeV while the pulling-down scenario is
favored. Stops can be light here, so Higgs sterility significantly
constrains them directly and indirectly except for blind spots. We
also study the LHC features of the whole stop sector facing a
sterile Higgs and find that, in virtue of decays between stops and
sbottom, characteristic signatures like same-sign leptons and multi
$b-$jets are promising probes.

\end{abstract}

\maketitle

\section{Introduction and motivations}

In the last two years, the ATLAS and CMS collaborations have
established the discovery of a new resonance, putatively the
long-sought standard model (SM)-like Higgs boson
$h$~\cite{LHC:Higgs}. It is a big milestone for the particle
physics. The more precise measurements on its particle properties
are still ongoing, but in light of the current
data~\cite{LHC:Higgs}, we know that it has a mass $m_h\simeq126$ GeV
(relatively heavy if interpreted in the minimal supersymmetric SM
(MSSM)), and moreover its main signatures are  consistent with
the SM predictions very well.

Actually, the highly SM-like Higgs boson emerges as data accumulating. In
the Higgs discovery, the channels with largest sensitivity are the
four lepton channel $h\ra ZZ^*\ra 4\ell$ and the di-photon channel
$h\ra \gamma\gamma$. The former does not show any significant
deviation from the SM prediction. While the latter, despite of
showing excess at the early stage, is steadily declining to the SM
case. The fermionic channels such as $h\ra b\bar b$ and $h\ra \tau
\bar \tau$ have smaller sensitivities, but the present hints of
these channels indicate that their signal strengthes are also within
the SM expectations~\cite{Mohr:2013eba}. Thereby, pessimistically
speaking, we may have to face a highly SM-like Higgs boson (dubbed
as sterile Higgs boson hereafter) in the near future. To quantify
Higgs sterility, we refer the LHC best experimental resolution which
is based on the 14 TeV LHC of 300 fb$^{-1}$, for
instance~\cite{Peskin:2012we}
\begin{align}\label{}
\frac{\Delta(\sigma_{\rm GF}{\rm Br}(2\gamma))}{\sigma_{\rm GF}{\rm
Br}(2\gamma)}:0.06,\quad \frac{\Delta(\sigma_{\rm GF}{\rm
Br}(ZZ))}{\sigma_{\rm GF}{\rm Br}(ZZ)}:0.09.
\end{align}
Resolution of ILC can be as good as 1$\%$, but the current numerical
tools can not match that. Thus, for main channels a deviation
$\lesssim$ 10$\%$ is a reasonable range of sterility.

As been well known, the Higgs signatures can be utilized to probe
new physics beyond the SM, e.g., the Higgs mixing with other states,
couplings to extra charged particles, and decaying into extra light
particles. As a matter of fact, all of them, especially the first
and second cases, occur in the supersymmetric SMs (SSMs). In the
SSMs, the SM Higgs sector is extended by another Higgs doublet like
in the minimal SSM (MSSM), and maybe one more singlet in the next to
MSSM (NMSSM)~\cite{Ellwanger:2009dp} (or
triplet~\cite{Kang:2013wm,Basak:2012bd}). Hence Higgs
doublet-doublet and doublet-singlet mixing (DSM) are expected.
Moreover, the stop sector, which significantly couples to $h$, has
effects on the Higgs mass and couplings as well. Therefore, it is of
importance to investigate implications of Higgs sterility on the
Higgs and stop sector. In this paper we analytically analyze the
feature of doublet-doublet mixing in the MSSM, and how it is
affected by DSM in the NMSSM. It is found that the doublet-doublet
mixing effect decouples as $1/M^2_A$ and $\tan\beta/M_A^2$,
respectively. Owing to $m_h$, in the MSSM the stop sector should be
heavy and is thus hardly constrained by Higgs sterility, except in
some limiting case. By contrast, in the NMSSM the whole stop sector
can be fairly light, so sterility acts. Besides, DSM can push-up or
pull-down $m_h$, with a degree bounded by Higgs sterility, as means
that the stop sector is also indirectly influenced by sterility.

With the resulted light stop ensemble which contains two stops and
light sbottom, we are interested in their LHC profiles.
They potentially provide a new angle on stop searches at the LHC. For
instance, generically speaking decays between stops and sbottom are
kinematically allowed and with large branching ratios, so a hard $W$
or $Z$ boson is produced. Taking into account the possible top quark
from the lightest stop decay, we thus expect signatures with same sign
leptons plus missing energy at the LHC. From our preliminary
analysis, this is a promising probe for the stop ensemble.

This paper is organized as follows. In Section II we investigate
implications of a sterile Higgs boson around 126 GeV on the Higgs
and stop sector, of the MSSM and NMSSM respectively. In the next
section an anatomy of the stop sector facing such a Higgs boson is
made. We analyze the decays of the stop ensemble and preliminarily
explore their characteristic signatures at the LHC. Discussion and
conclusion are casted in Section IV and some necessary and
complementary details are given in the Appendices.

\section{Implications of a sterile Higgs boson in the MSSM and NMSSM}

The current data may point to a Higgs boson with highly SM-like
couplings, so seemingly it does not convey much information of new
physics to us. Such a sterile Higgs boson places stringent bounds on
Higgs couplings which, in the SSMs, tend to show deviations from the
SM predication. In this section, taking the MSSM and NMSSM as
examples, we investigate implications of Higgs sterility on the
Higgs sector, which exhibits Higgs mixings, and on the stop sector,
which has a notable effect on both mass and couplings of the Higgs
boson. Numerical study is employed as well.

\subsection{A sterile Higgs boson in the MSSM}

\begin{figure}[htb]
\includegraphics[width=3.2in]{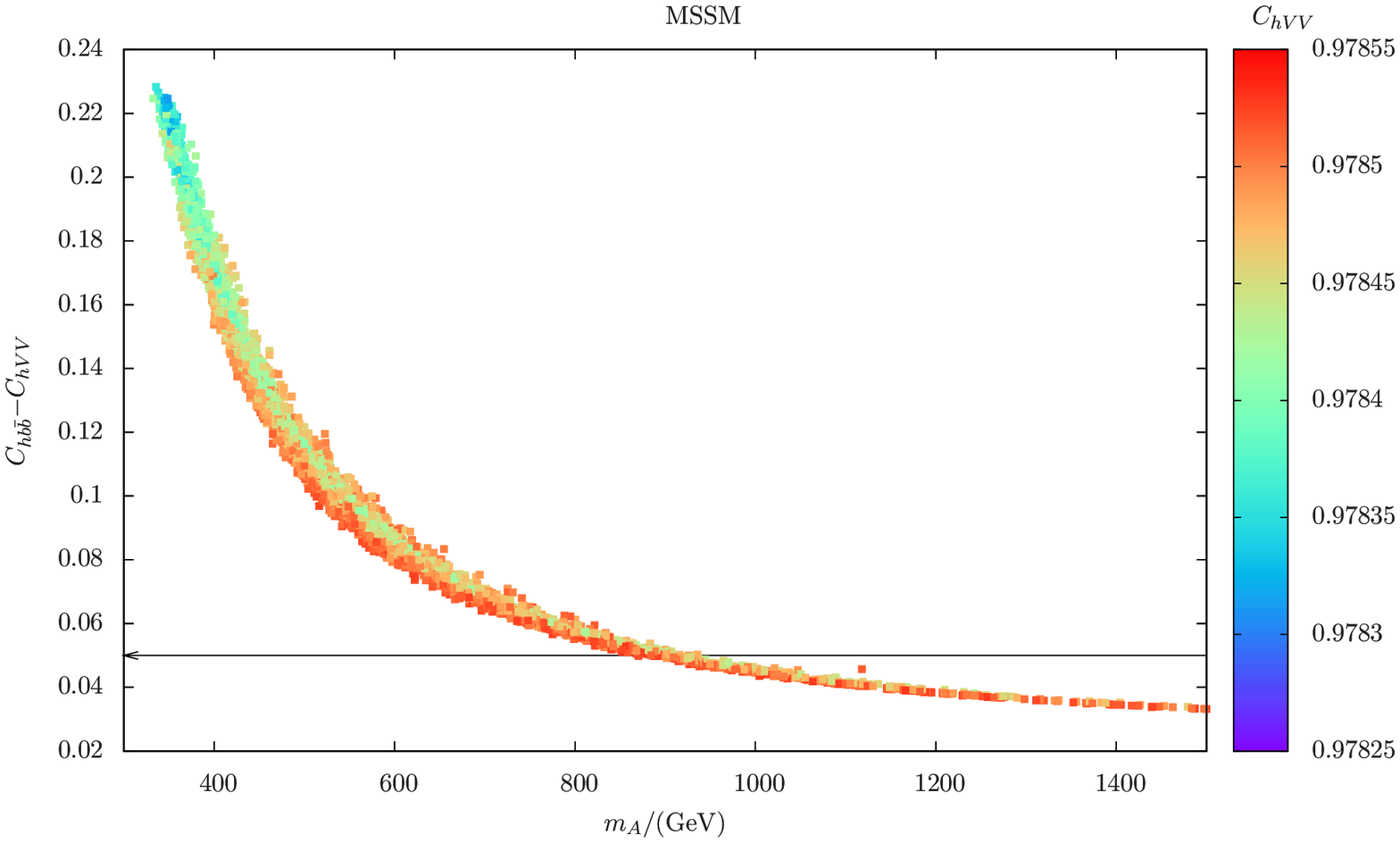}
\includegraphics[width=3.2in]{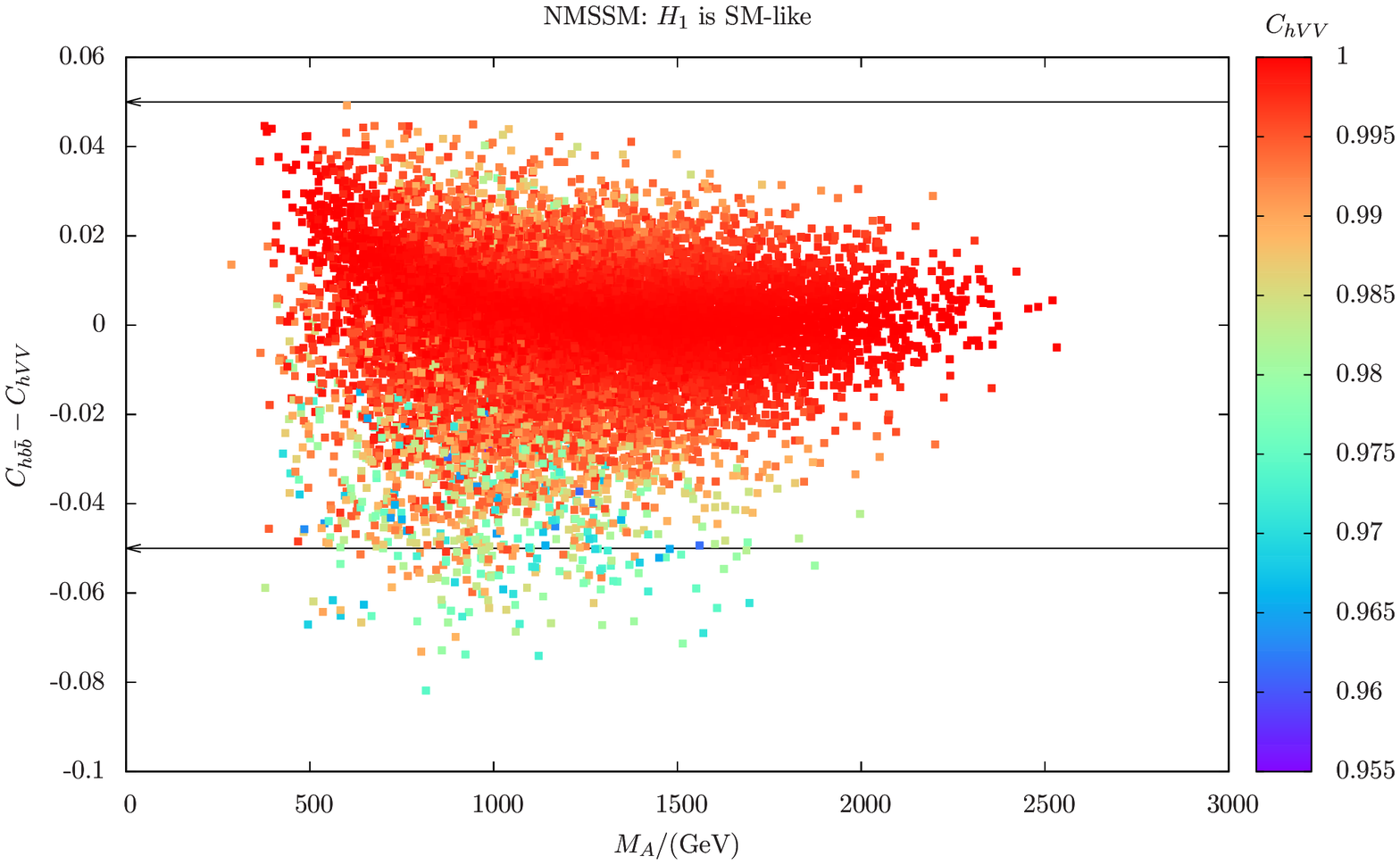}
\includegraphics[width=3.2in]{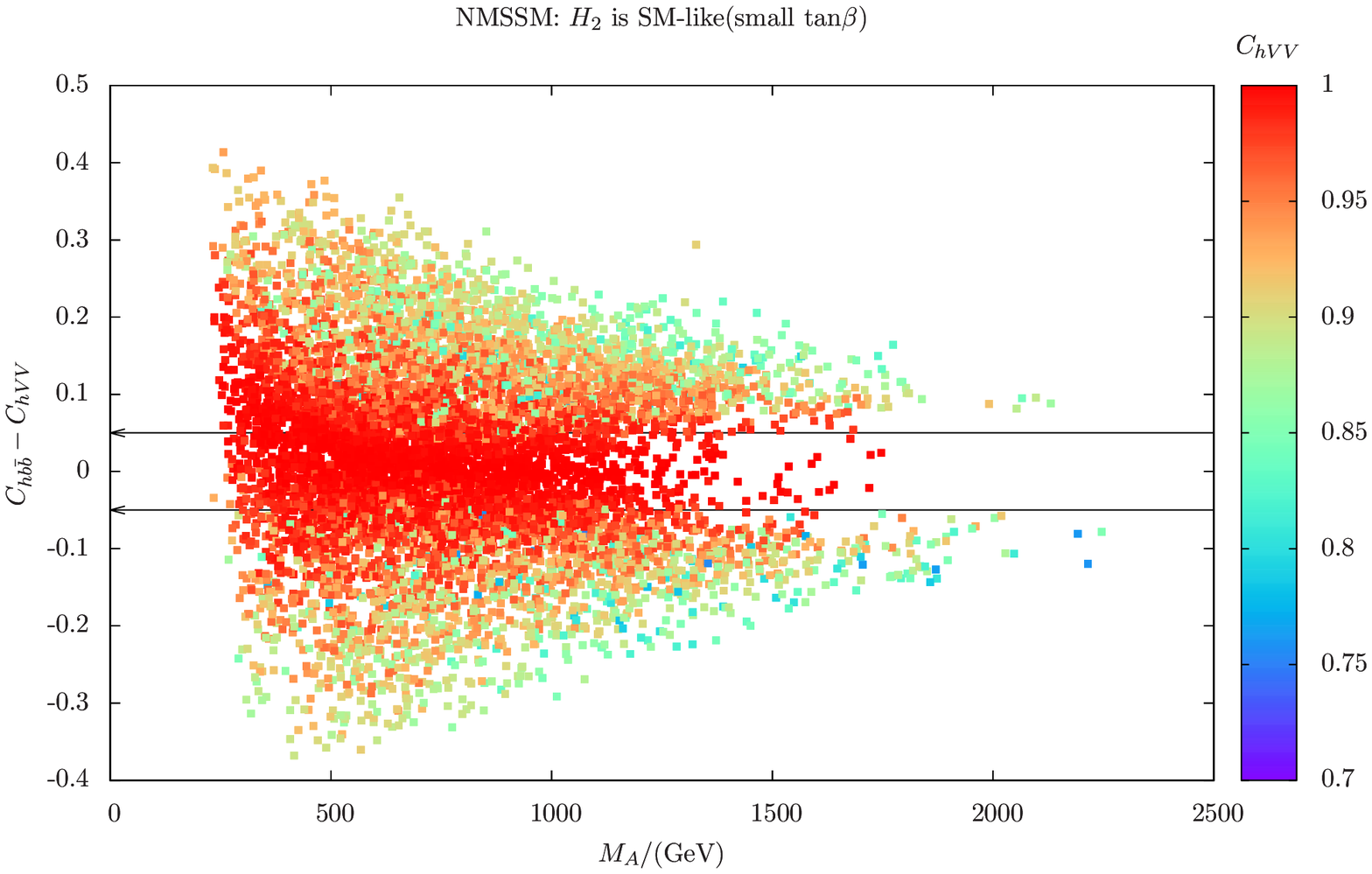}
\includegraphics[width=3.2in]{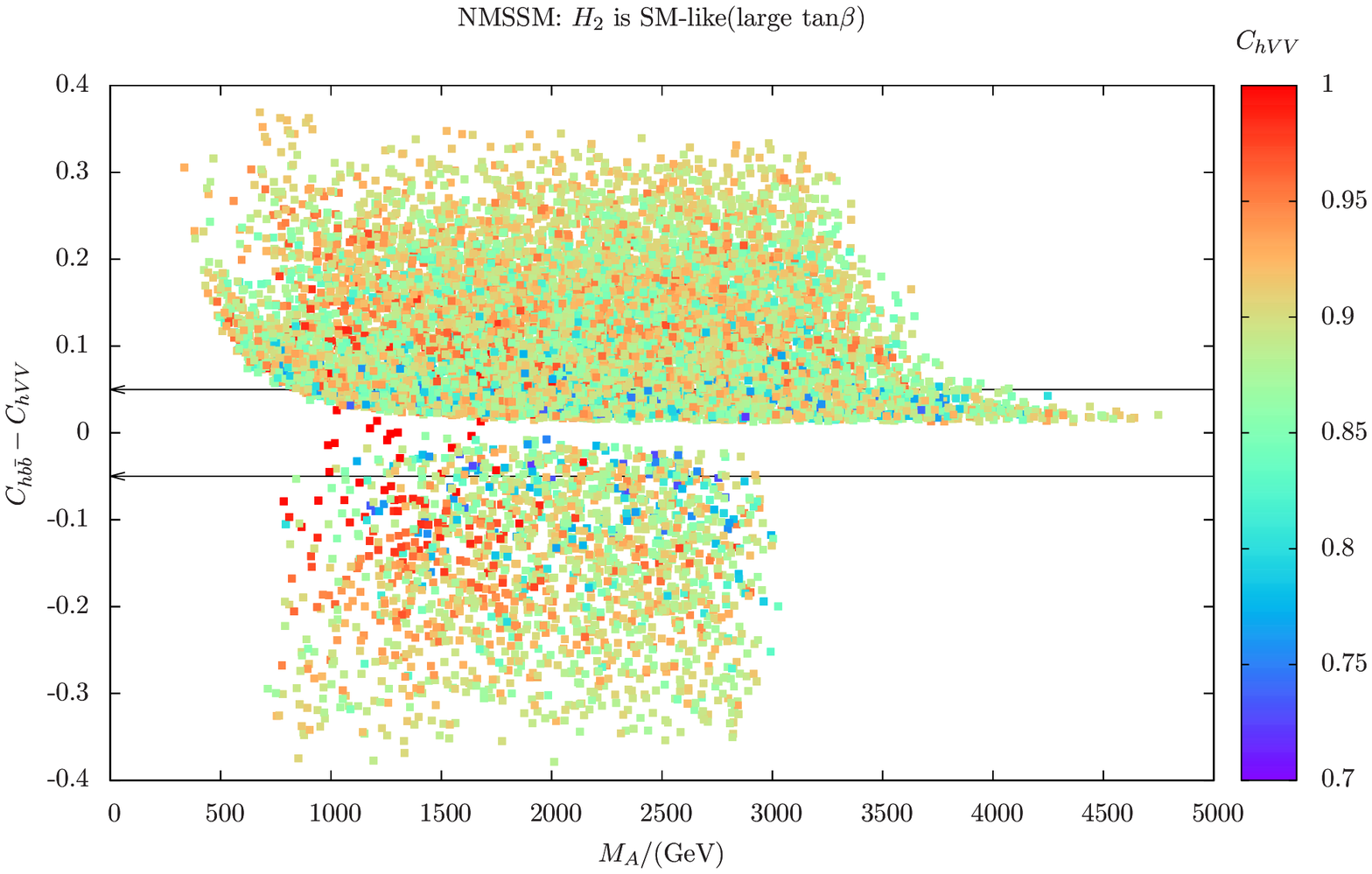}
\caption{\label{fig1} Inspecting Higgs sterility on the $\wt
C_{hb\bar b}-M_A$ plane, with color code denoting $C_{hVV}$. The
reduced couplings with a tilde are subtracted by $C_{hVV}$ so as to
isolate the universal mixing effect. Top left: MSSM; Top Right:
NMSSM in the puling-down scenario; Bottom left/right: NMSSM in the
pushing-up scenario with a small/large $\tan\beta$.}
\end{figure}

In the MSSM we have two Higgs doublets $H_u$ and $H_d$. The mixing
effects between them are not difficult to be analyzed. They lead to
the tree-level reduced couplings of the SM-like Higgs boson (All
notations are casted in Appendix~\ref{convention}.):
\begin{align}\label{reduce}
C_V=\sin(\beta-\alpha),\quad C_t=\f{\cos\alpha}{\sin\beta},\quad
C_{hb\bar b}=-\f{\sin\alpha}{\cos\beta},
  \end{align}
with $\tan\beta=v_u/v_d$. The mixing angle between the heavy and
light (SM-like) CP-even Higgs boson $\alpha$ is given
by~\cite{Djouadi:2005gj}
\begin{align}\label{reduce}
-\pi/2\leq\alpha=\f{1}{2}\arctan\left[\tan2\beta(M_A^2+m_Z^2)/(M_A^2-m_Z^2)
\right] \leq0.
  \end{align}
In the nearly decoupling region $M_A^2\gg m_Z^2$ and
$\tan\beta\gg1$, the expressions in Eq.~(\ref{reduce}) are
approximated to be~\cite{Djouadi:2005gj}
\begin{align}\label{reduce}
C_V\ra 1-\f{2m_Z^4}{M_A^4\tan^2\beta},\quad
C_t=1-\f{2m_Z^2}{M_A^2\tan^2\beta},\quad C_{hb\bar b}=1+\f{2m_Z^2}{M_A^2}.
  \end{align}
As one can see, only $C_{hb\bar b}$ can be appreciably affected,
concretely speaking, enhanced. In that case, the MSSM predicts a
universal suppression of the signature strengths except for these
involving $b\bar b$ which should be close to unit~\footnote{Such a
prediction is of great importance to find a smoking gun for the
exotic Higgs bosons in the MSSM. As far as our knowledge, this point
is not explicitly pointed out by any reference, despite of a
relevant study~\cite{Arbey:2013jla}. We leave a specific study about
it elsewhere.}. From the first panel of Fig.~\ref{fig1} it is seen
that, to meet Higgs sterility we need to set $M_A \gtrsim$ 900 GeV.
This is obviously heavier than the tree-level estimation $\gtrsim
600$ GeV, owing to the radiative correction which is enhanced by a
large $\tan\beta$~\cite{Djouadi:2005gj}.

We now turn to the implication of Higgs sterility on the stop
sector. It is well known, due to the significant coupling to $H_u$,
the stop sector plays a crucial role in moulding properties of the
SM-like Higgs boson. Firstly, it is related to origins of the Higgs
boson mass. In SSMs the Higgs boson mass can be expressed as
\begin{align}\label{Hmass}
m_h^2= \L m_Z^2  \cos^22\beta+\Delta m_h^2\R+ m_Z^2 f(\ld,\beta).
  \end{align}
The first term is predicted by the MSSM. It consists of the
tree-level contribution from Higgs quartic term, determined by
D-terms, as well as the stop radiative correction
\begin{align}
\Delta m_h^2=\frac{3m_t^4}{4\pi^2v^2}\left[\log{\frac{m_{\wt
t}^2}{m_t^2}} +\frac{X_t^2}{m_{\wt t}^2}\L1-\frac{X_t^2}{12m_{\wt
t}^2}\R\right]~,~\,
\end{align}
with the average stop mass $m_{\wt t}=\sqrt{m_{\tilde{t}_1}
m_{\tilde{t}_2}}$ (Stop parameters are defined in
Appendix~\ref{MH2}). The second term denotes contributions from
extra tree-level Higgs quartic terms such as in the NMSSM discussed
below. Secondly, stops, which carry both QCD and QED charges, modify
the Higgs effective couplings to gluons and photons, e.g., by a
shift in the Higgs-gluon reduced
coupling~\cite{Carmi:2012in,Ajaib:2012eb}
 \begin{align}\label{rGG}
\delta C_{hGG}=\f{\delta r_g}{r_{{\rm SM},g}}\approx
1+\f{1}{4}\L\f{m_t^2}{m_{\wt t_1}^2}+\f{m_t^2}{m_{\wt
t_2}^2}-\f{X_t^2}{m_{\wt t}^2} \f{m_{t}^2}{m_{\wt t}^2}\R.
  \end{align}
The convention can be found in Appendix.~\ref{convention}.
Therefore, with light stops or/and large stop mixing, Higgs
sterility may be violated.

In the MSSM almost half of $m_h$ origins from the stop radiative
correction. To achieve a large $\Delta m_h^2$ and keep stops as
light as possible at the same time, we have to rely on a large stop
mixing, says in the stop maximal mixing scenario with $X_t^2\simeq
6m_{\wt t}^2$. Light stops are chased after for the sake of both
naturalness and their detection at the LHC. Then Higgs sterility
excludes a part of the parameter space of light stops. We would like
to stress that, \emph{light stops and large stop mixing may result
in a substantial cancelation between terms in the bracket of
Eq.~(\ref{rGG}), so a blind spot exists in Higgs sterility}. In
other words, light stops may hide behind the sterile Higgs boson. It
is straightforward to derive the condition for that:
\begin{align}\label{}
m_{\wt t_1}^2+m_{\wt t_2}^2=m_{LL}^2+m_{RR}^2=X_t^2.
  \end{align}
The top left panel of Fig.~\ref{fig2} shows that Higgs sterility is
absolutely null and void. However, it is not always the case. It is
blamed to our parameter setting for the stop sector shown in
Eq.~(\ref{push:R}), which just drives the light stop around 350 GeV
into the blind spots. In principle, one light stop is allowed to be
rather light if we set another stop very heavy.

\subsection{A sterile Higgs boson in the NMSSM}

The Higgs sector of the NMSSM is further extended by a singlet $S$,
which dramatically changes the Higgs phenomenologies. Above all, it
is able to enhance $m_h\simeq126$ GeV without turning to heavy stops
and thus is regarded as a benchmark model for natural
SUSY~\cite{Kang:2012sy}. The Higgs sector of the model, in the scale
invariant form, is given by
\begin{align}\label{WS}
W\supset&\ld S H_u\cdot H_d+\f{\kappa}{3}S^3,\cr
 -{\cal L}_{soft}\supset & \ld A_{\ld}
SH_u\cdot  H_d+\f{{\kappa}}{3}A_{\kappa} S^3+h.c.
  \end{align}
There are three CP-even Higgs bosons out of this Higgs sector. To
understand Higgs mixing and mass, it is convenient to work in a
basis defined as~\cite{PQ:mass,Kang:2013rj}
{\small\begin{align}\label{} H_u^0=&v_u+\f{1}{\sqrt{2}}\L
S_1\cos\beta+S_2\sin\beta\R,~ H_d^0=v_d+\f{1}{\sqrt{2}}\L
S_1\sin\beta-S_2\cos\beta\R,~ H_S= v_s+\f{S_3}{\sqrt{2}},
  \end{align}}The mass eigenstates
$H_{i=1,2,3}$ (masses in ascending order) are related with $S_i$ via
$O$, which is defined through $OM_{S}^2O^T={\rm
Diag}(m_{H_3}^2,\,m_{H_2}^2,\,m_{H_1}^2)$ with $M_{S}^2$ the Higgs
mass square matrix in the basis defined above (entries of $M_{S}^2$
see Appendix.~\ref{MH2}). Neglecting mixing effects, the tree-level
$m_h$ is nothing but $(M_S^2)_{22}$ which is a function of $\ld$ and
$\tan\beta$, namely in Eq.~(\ref{Hmass})
\begin{align}\label{}
f={\ld^2}\sin^22\beta/{g^2}.
  \end{align}
Plotting the contour of $m_h$ on the $\tan\beta-\ld$ plane, $\ld=\L
g_1^2+g_2^2\R^{1/2}\approx0.53$ is a critical line (Along it $m_h$
independes on $\tan\beta$.): For $\ld>0.53$, the large $\ld-$effect
is working, and lowering $\tan\beta$ helps to enhance $m_{h}(>m_Z)$;
While for $\ld<0.53$ the situation is opposite. But the
doublet-singlet mixing (DSM) effect modifies $m_h$, which will be
discussed soon later.

With DSM, studying features of Higgs signature in the NMSSM is much
more complicated than that of the MSSM (See some related
works~\cite{Choi:2012he,Badziak:2013bda}). But we find that for our
purpose, the main features can be manifested by means of a simple
approximate method. For definiteness, we focus on $h=H_2$ and the
case with $h=H_1$ can be discussed similarly. Then the reduced
couplings of $H_2$ at tree-level are calculated to be
\begin{align}\label{C:tree}
C_{2,V}=O_{22},\quad C_{2,t}\simeq O_{22}+O_{21}\cot\beta,\quad
C_{2,b}=O_{22}-O_{21}\tan\beta.
\end{align}
In most cases, $O_{21}\cot\beta\ll1$ can be safely neglected, and
thus we get the universal reduction factor $C_{2,V}\approx
C_{2,t}=O_{22}<1$, which is mostly ascribed to DSM. The
doublet-doublet mixing along with the DSM violate that universality
by allowing a widely varied $C_{2,b}$. Moreover, it is noticed that
as opposed to that of the MSSM, here $C_{2,b}$ can be either larger
or smaller than unit. To see this, we make use of the equation
$O_{1i}(M_S^2)_{ij}O_{2j}=0$ to find out $O_{21}\tan\beta$ at the
leading order:
\begin{align}\label{DSM}
-O_{21}\tan\beta\simeq -\sin^2\beta\cos2\beta\f{2\L
m_Z^2-\ld^2v^2\R}{M_A^2}-\tan\beta\f{O_{13}
(M_S^2)_{23}+O_{23}(M_S^2)_{13}}{M_A^2}.
\end{align}
This simple formula reveals the impact of DSM. When $\tan\beta\gg1$,
the first term reproduces its corresponding expression given in
Eq.~(\ref{reduce}), up to the replacement $m_Z^2\ra m_Z^2-\ld^2v^2$.
Thus, given $\ld\gtrsim 0.6$ this term becomes negative. It is one
of the difference between the MSSM and NMSSM, but is attributed to
the new quartic term rather than DSM. The DSM effect is encoded in
the second term of Eq.~(\ref{DSM}). One can find that, when we have
a small $\tan\beta\sim 1$ and moreover a properly light $M_A$ (not
as light as the one considered in the
Ref.~\cite{Christensen:2013dra}), the second term tends to be
dominant. But it has an indefinite sign, and consequently $\Gamma
(h\ra b\bar b)$ may be either increased or decreased. To show how
does the DSM effect change Higgs signatures, we give the signature
strength of Higgs to di-photon in the gluon fusion channel:
\begin{align}
R_{gg}^{H_2}(\gamma\gamma)\approx &O_{22}^2\f{1+2\L\delta
r_{2,g}+\delta r_{2,\gamma}/r_{{\rm SM},\gamma}\R/O_{22}}
{1-1.17\tan\beta\,{O_{21}}/{O_{22}}+0.18\,\delta r_{2,g}/O_{22}},
\end{align}
where the stop contributions have been formally took into account.
In summary, mixings in the NMSSM and MSSM are similar in the sense
of the importance of $C_{hb\bar b}$, however, their quantitative
consequences are noticeably different. In particular, \emph{the DSM
effect in $O_{21}\tan\beta$ is enhanced by a large $\tan\beta$,
which makes it decouple not as $1/M_A^2$ but as $\tan\beta/M_A^2$}.
As a result, it may be still significant even for $M_A \gtrsim 3$
TeV, see the bottom-right panel of Fig.~\ref{fig1}.
\begin{figure}[htb]
\includegraphics[width=3.2in]{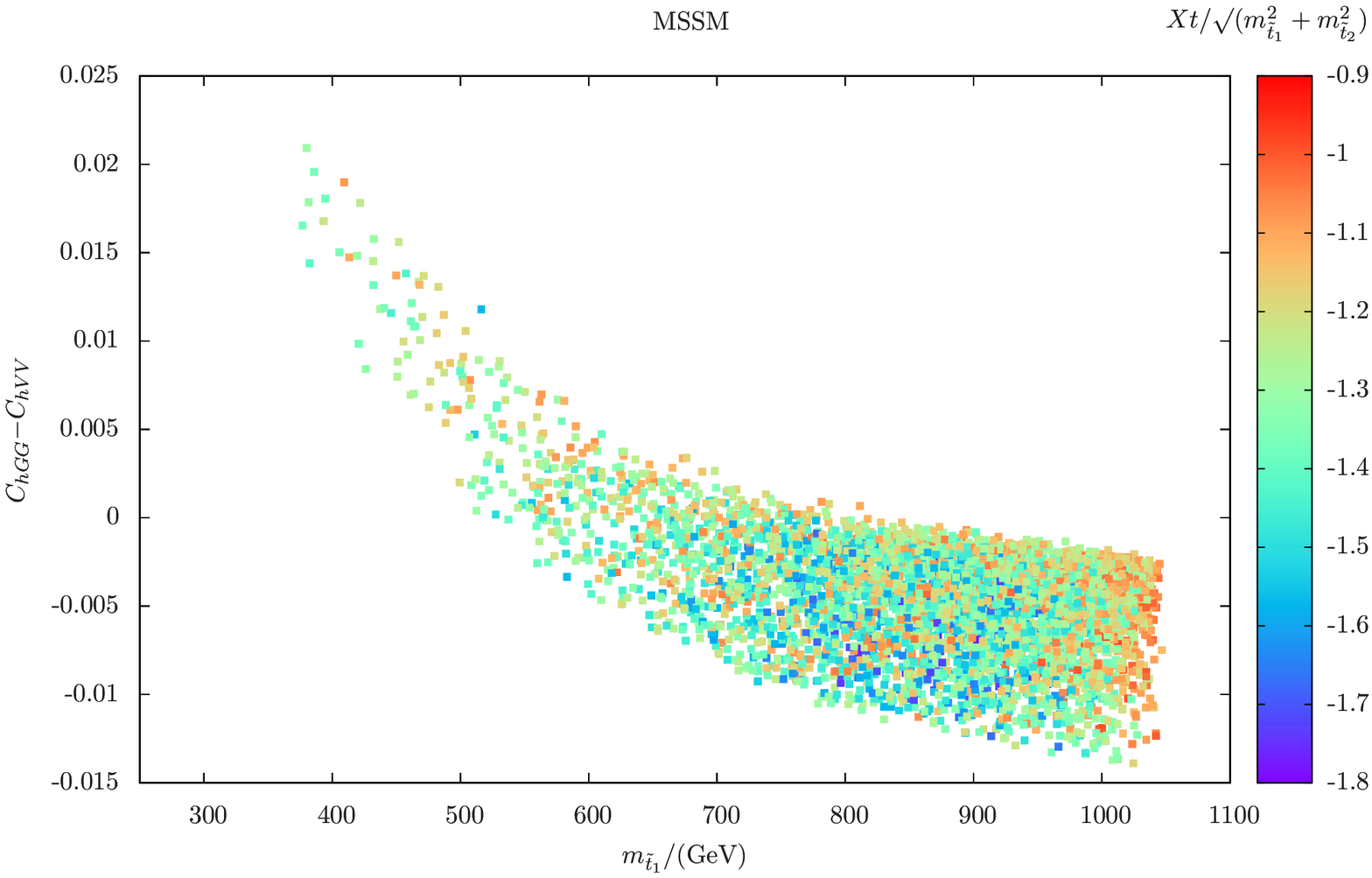}
\includegraphics[width=3.2in]{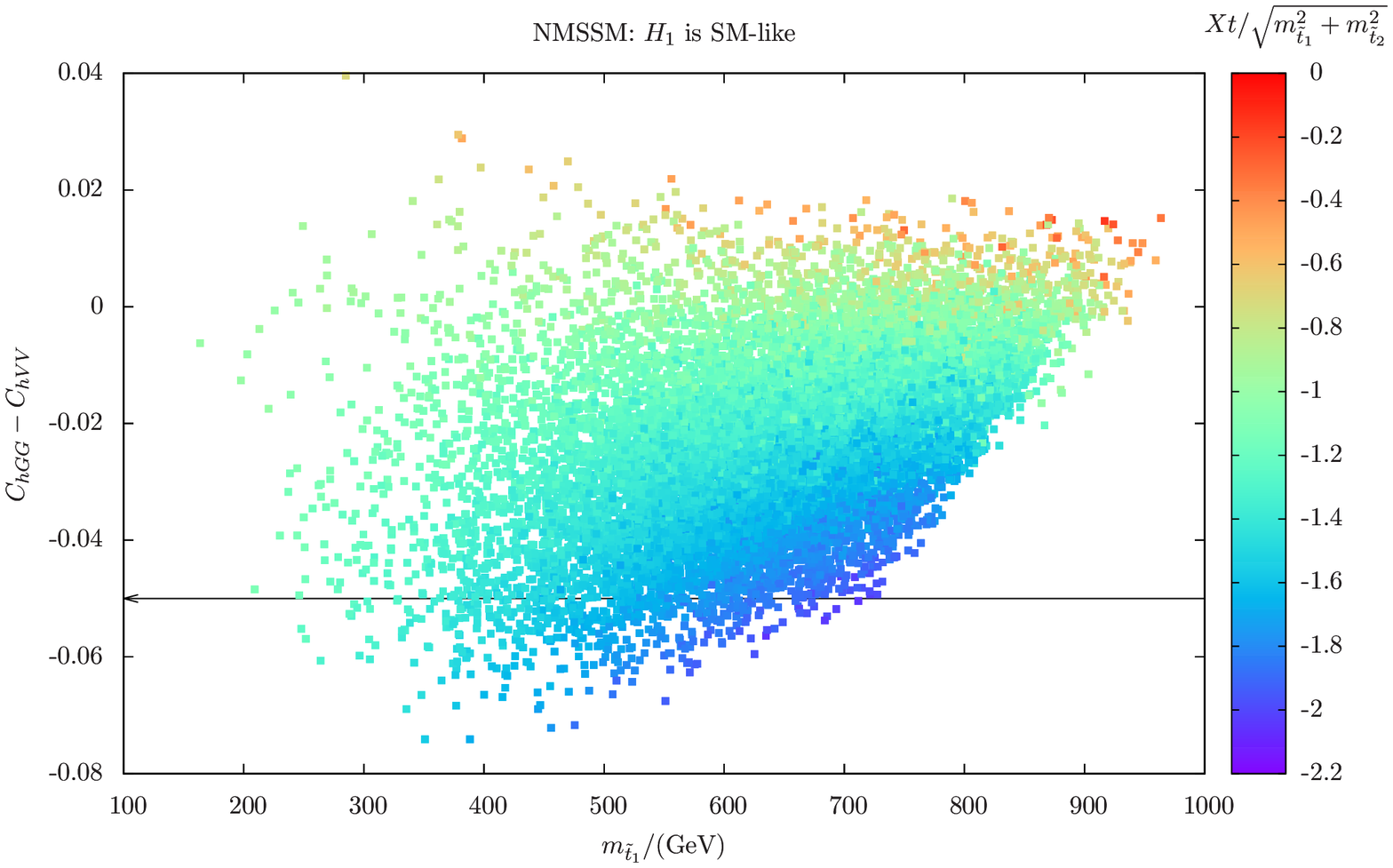}
\includegraphics[width=3.2in]{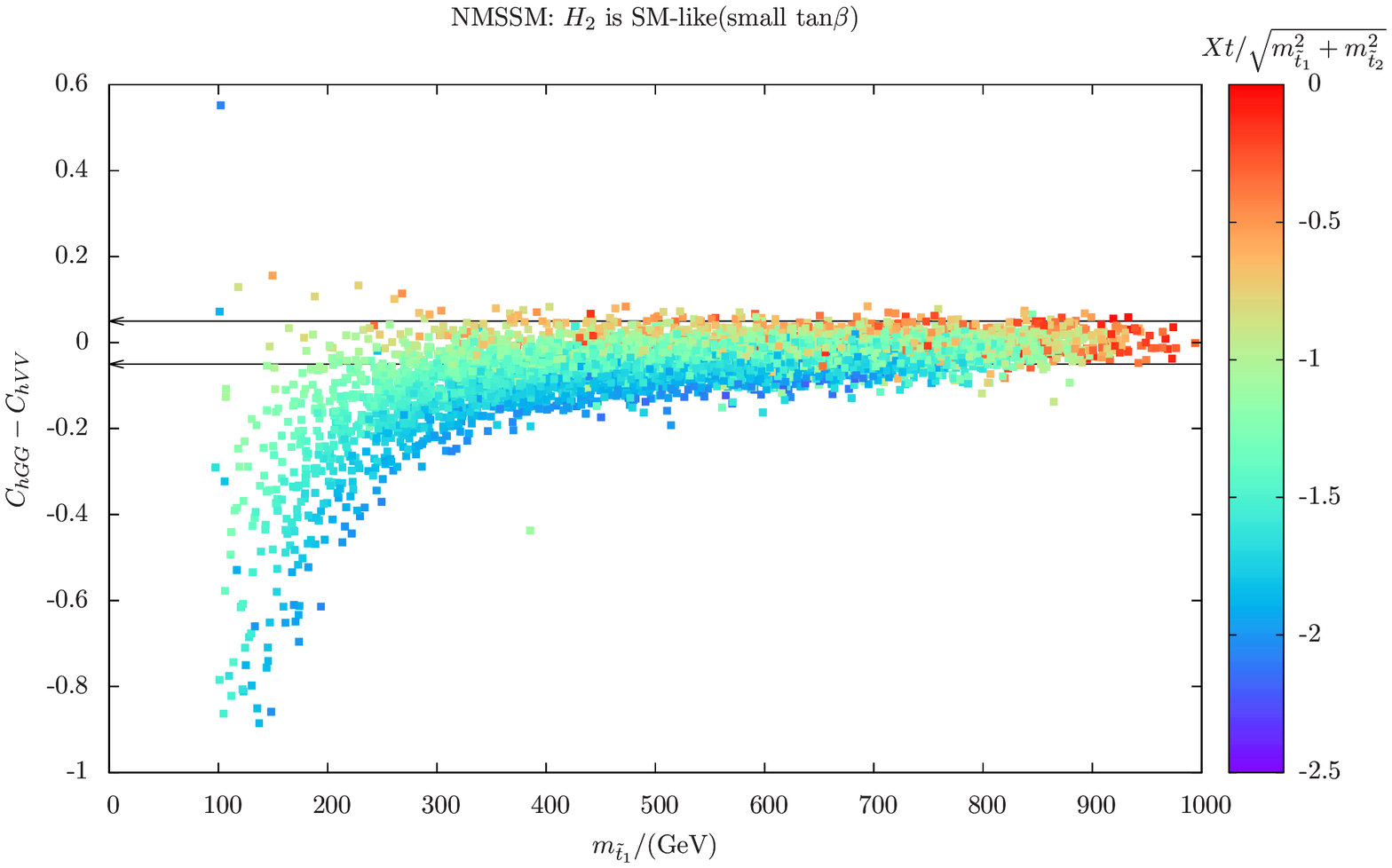}
\includegraphics[width=3.2in]{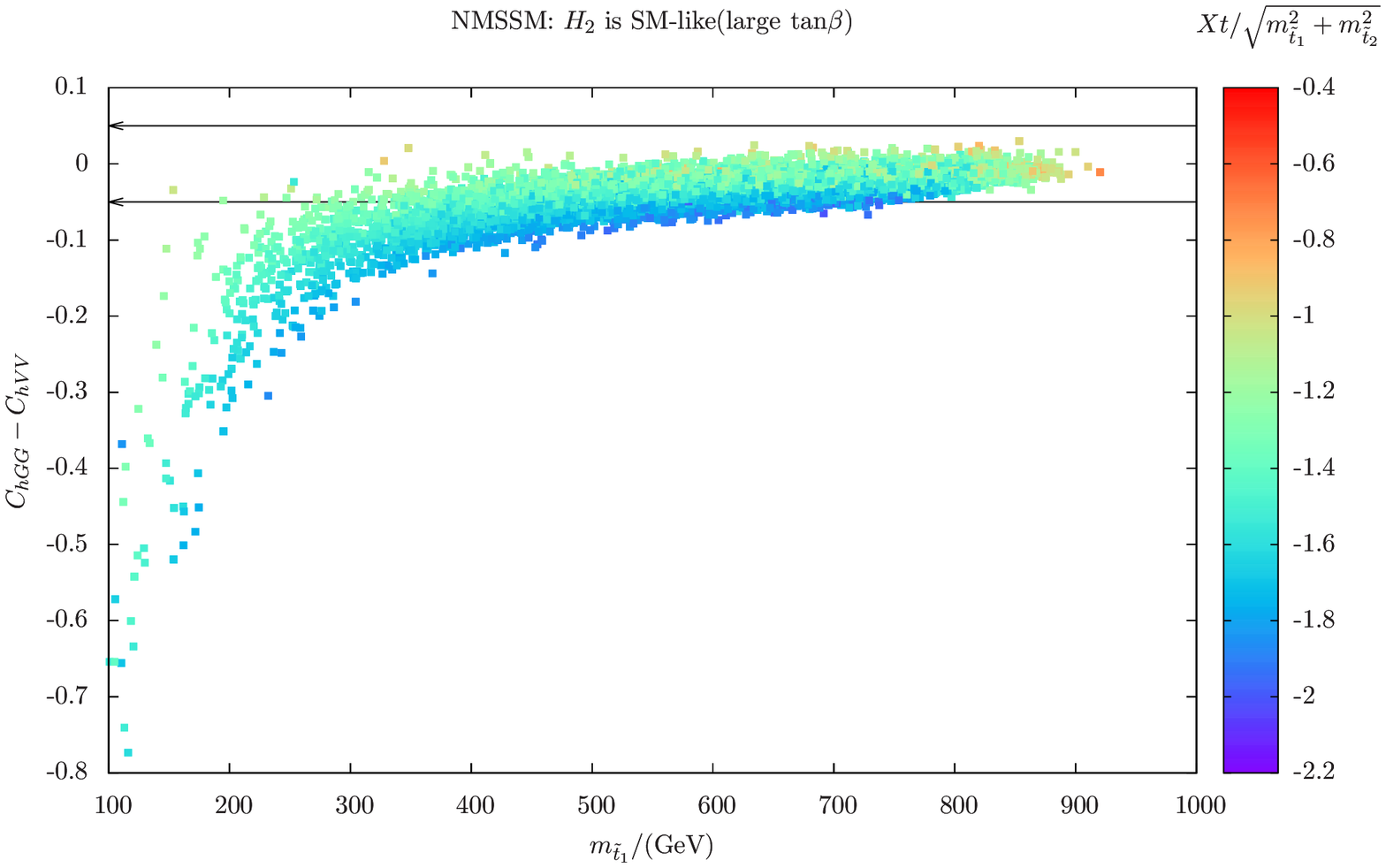}
\caption{\label{fig2} Inspecting Higgs sterility on the $\wt
C_{hGG}-m_{\wt t_1}$ plane, with color code denoting the quantity
$X_t/\sqrt{m^2_{\wt t_1}+m^2_{\wt t_2}}$ which reflects blindness of
sterility. Figures are ordered the same with Fig.~\ref{fig1}.}
\end{figure}

DSM influences not only the Higgs couplings but also Higgs mass.
Concretely, the DSM effect pushes-up or pulls-down
$m_h$~\cite{Kang:2012sy,Chang,Cao:2012fz}, depending on $H_2=h$ or
$H_1=h$. In what follows we will investigate implications of Higgs
sterility on each scenario, respectively.
\begin{description}
  \item[Revisit to the pushing-up scenario facing a sterile Higgs]
  We first consider $H_2=h$, namely the pushing-up scenario which
  is characterized by an even lighter (than $H_2$) CP-even Higgs boson
 $H_1$~\footnote{How to detect this light Higgs boson,
 or more broadly speaking the extra light Higgs states predicted in the NMSSM,
 is challenging but interesting~\cite{Kang:2013rj,Christensen:2013dra,Cerdeno:2013cz}.}.
Realization of this scenario is important. First of all, it
requires $(M_{S}^2)_{22}$ $>$ $(M_{S}^2)_{33}$. From
Eq.~(\ref{Higgs:even1}) it is seen that a moderately small $\mu$ and
not too large $\kappa/\ld$ are favored to make $(M_{S}^2)_{33}$
sufficiently small. Furthermore, a properly large doublet-singlet
mixing term (${M_S^2})_{23}$~\cite{Kang:2012sy} is needed: On the
one hand, it should be large enough to guarantee a sizable $\Delta
m_h$; On the other hand, it should be small enough to prevent a
tachyon. Then typically we need
\begin{align}\label{matr}
(M_S^2)_{23}= 2\ld \mu v
\left[1-\L\f{A_\ld}{2\mu}+\f{\kappa}{\ld}\R\sin2\beta\right]
\sim{\cal O}(1000){\rm\,GeV^2},
\end{align}
except very degenerate $(M_{S}^2)_{22}$ and $(M_{S}^2)_{33}$.
Thereby, the region with $\ld\sim1$, $\tan\beta\sim1$ and
$\mu\sim200$ GeV accords well with the pushing-up scenario.
Actually, this region takes full advantage of NMSSM effects to
enhance $m_h$ and is extensively
studied~\cite{Kang:2012sy,Ellwanger,Cao:2012fz}. But even for a
larger $\tan\beta$ and/or smaller $\ld$, one can still turn to a
large (but not exceedingly large) $A_\ld$ to compensate their
suppression on $(M_S^2)_{23}$ and thus give a sizable
pushing-up effect~\footnote{Such a scenario was briefly discussed in
Ref.~\cite{Kang:2012sy} and then numerically studied by
Ref.~\cite{Aparicio:2012vk,Badziak:2013bda}. Here we present a more
detailed numerical analysis.}. The right panel of Fig.~\ref{fig3}
confirms the analysis.

We are at the position to quantify the pushing-up effect.
Ref.~\cite{Kang:2012sy} took an approximate method. It starts from
the previously defined basis, in which the doublet sector has been
approximately diagonalized, with two eigenvalues $(M_S^2)_{22}$ and
$(M_S^2)_{11}(>(M_S^2)_{22})$ and the lighter state being the
dominant component of $h$. It decouples the heavier state and
discusses the DSM effect in the latter $2\times 2$ submatrix of
$M_S^2$. This treatment neglects other DSM effects, which may be
important especially in the region with a relatively small $M_A$ and
$\tan\beta$. In this work we instead use a numerical method. We
diagonalise first the doublet sector then the full mass matrix, and
each time get the SM-like Higgs boson mass $m_{h'}$ and $m_{h}$,
respectively. Then the DSM pushing-up effect can be measured by
\begin{align}\label{}
\Delta m_h\equiv m_{h}-m_{h'},
\end{align}
which is the exact result, including all DSM effects. Since the
amount of pushing-up, $\Delta m_h$, is related to DSM, a sterile
Higgs boson raises doubts about it. With numerical results we will
find that, after imposing Higgs sterility (and the LEP upper
bound~\cite{Schael:2006cr} on $H_1$ as well), the resulted
pushing-up effect is indeed mild, typically $\Delta m_h\lesssim$ 5
GeV. This can be clearly seen in Fig.~\ref{fig3}.

\begin{figure}[htb]
\includegraphics[width=3.2in]{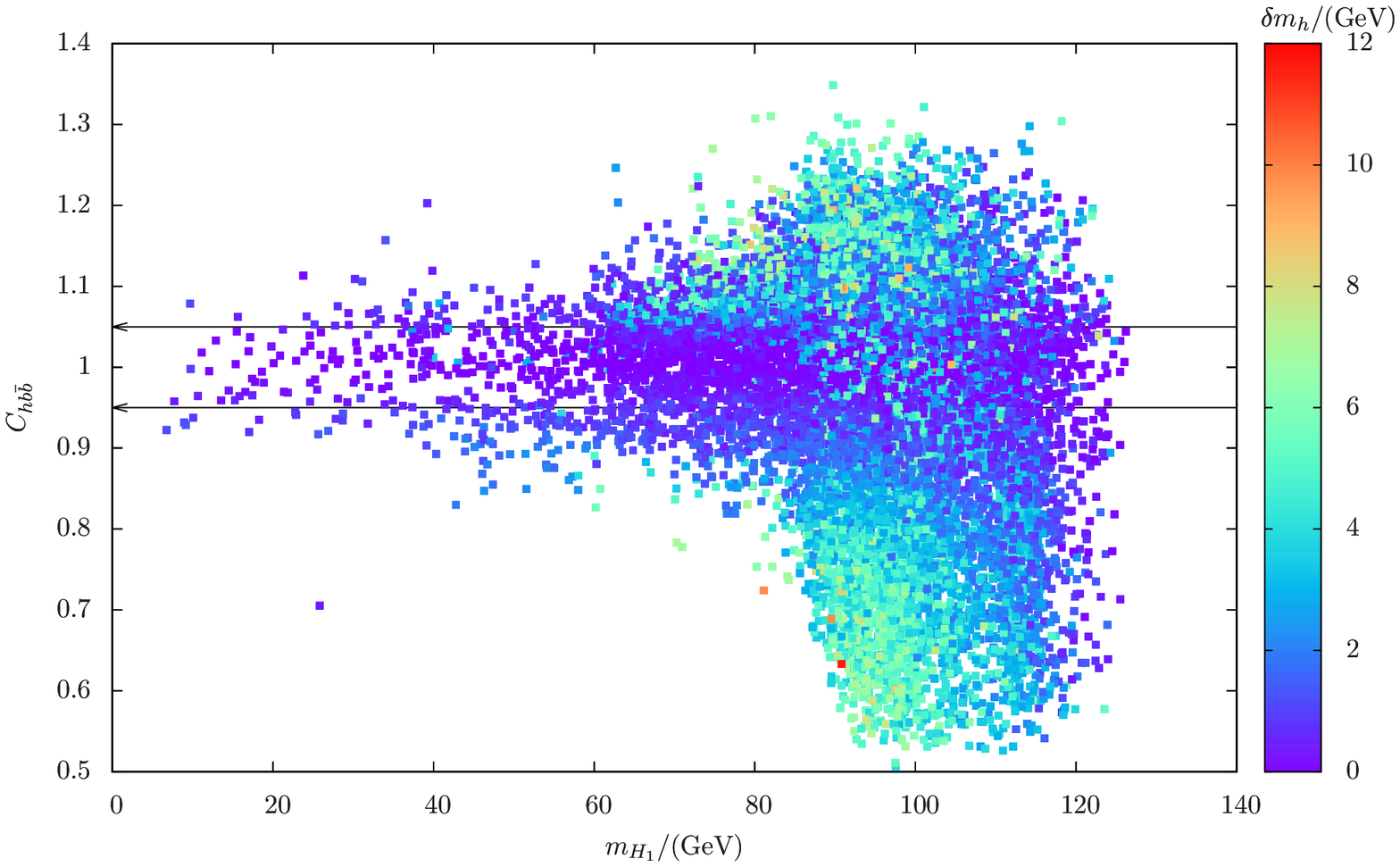}
\includegraphics[width=3.2in]{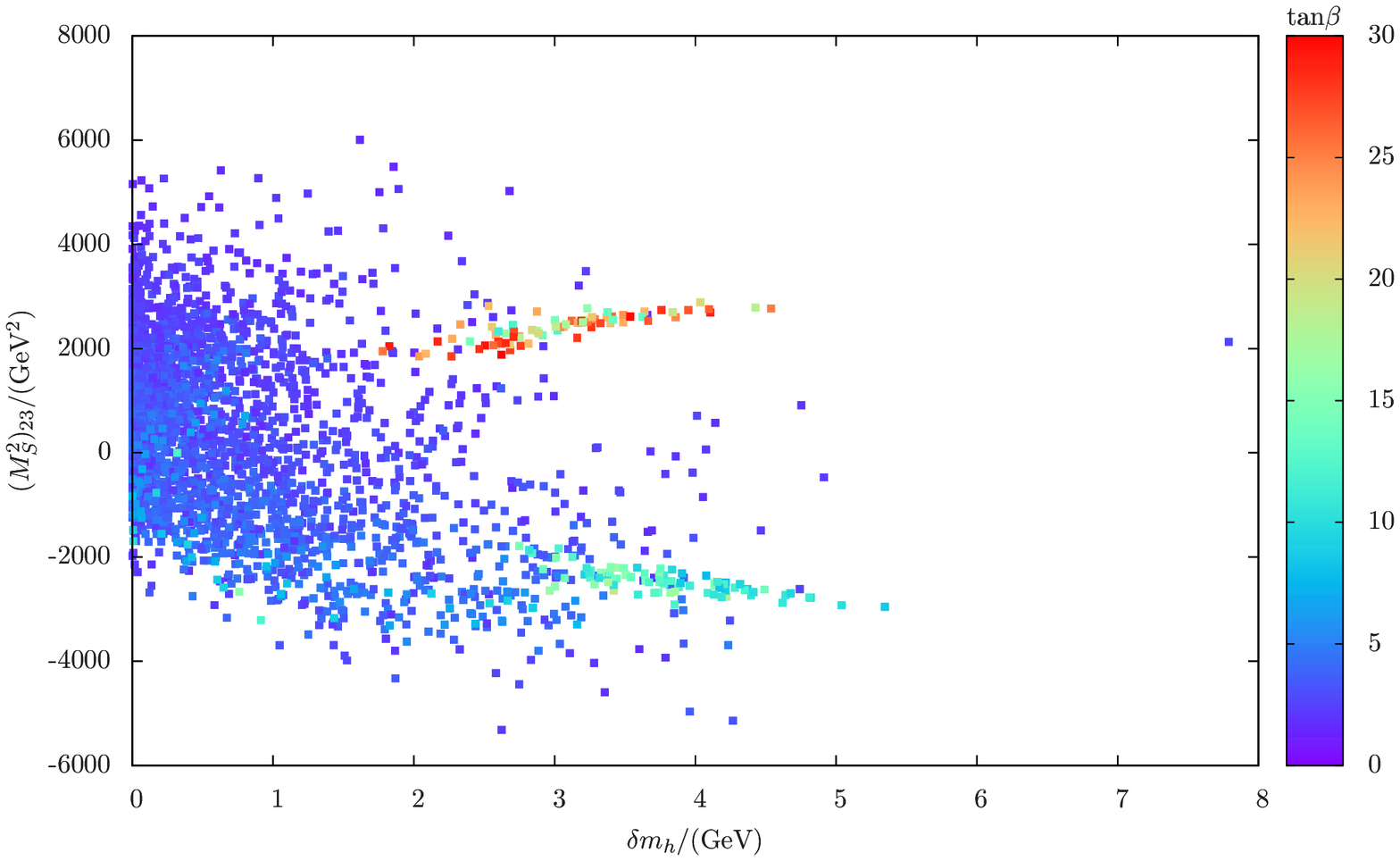}
\caption{\label{fig3} Left panel: Distribution of $\delta m_h$ on
the $\wt C_{hb\bar b}-m_{H_1}$ plane; Right panel: $\delta m_h$
versus doublet-singlet mixing element $(M_S^2)_{23}$, with color
code denoting $\tan\beta$.}
\end{figure}

\item[Is the pulling-down region favored?]
  We now turn our attention to the case $(M_S^2)_{22}<(M_S^2)_{33}$.
Then $h=H_1$ and we confront with the pulling-down effect. The
reduced couplings $C_{1,X}$ can be derived analogue to $C_{2,X}$. To
weaken the pulling-down effect to the most extent, one generically
expects a smaller DSM, which implies a suppressed DSM effect on
$C_{1,b}$ (more precisely, $O_{31}\tan\beta$). Moreover, compared to
the MSSM, in this scenario the doublet-doublet mixing effect $\simeq
2m_Z^2/M_A^2$ is also considerably attenuated, by the new large
quartic term $\ld^2v^2$ and a small $\tan\beta$ as well (See the
first term of Eq.~(\ref{DSM})). Therefore, $O_{31}\tan\beta$ is
slight and becomes slighter as $m_h$ becomes heavier. This explains
why in the pulling-down scenario the Higgs di-photon excess for a
126 GeV Higgs boson is not significant~\cite{Kang:2012sy}. However,
viewing from Higgs sterility, this scenario is favored. It is
manifest in the top right of Fig.~\ref{fig2}, where Higgs sterility
is almost automatically implemented. In addition, in this scenario
the LEP bound does not concern us.

If $\ld\ll1$, we essentially go to the MSSM limit, which has been
discussed above. So we only consider the large $\ld$ and small
$\tan\beta$ case, which retains the $\ld-$effect to enhance $m_h$
and hence we do not badly need heavy stops.

\end{description}

Since the NMSSM readily accommodates a light stop sector, direct
constraints from Higgs sterility is powerful here. Recalling that
DSM has effects on $m_h$, thus Higgs sterility is able to indirectly
constrain the stop sector. This kind of constraint is most
remarkable in the region where the $\ld-$effect is moderate or even
negligible and then we rely on the pushing-up effect and stop
radiative correction. To check that we compare the pushing-up
scenario with a large $\tan\beta$ and small $\tan\beta$ (see
Fig.~\ref{fig2}): Before imposing the Higgs sterility bound, both
cases allow a light stop $\sim100$ GeV, but imposing the bound
(largely) excludes $m_{\wt t_1}<250$ GeV and 150 GeV, respectively.
Note that as explained before, a light stop may lie in the blind
spot of Higgs sterility and thus is not excluded.

To end up this subsection, we would like to make a comment on the
relationships among the DSM effect, its modification on the
signatures and mass of Higgs boson. A significant DSM effect is
reflected in $O_{22}$ which shows a deviation from 1, as well as in
$O_{23,32}$ which should be relatively large. Generically, it would
lead to an universal suppression of Higgs signature strengths, by
$O_{22}^2$. However, in some cases the DSM effect, as shown
previously, can distort $C_{2b}$ such that the total decay width of
Higgs boson decreases substantially, and then some of strengths such
as di-photon rate are enhanced~\cite{Ellwanger,Cao:2012fz}. But such
kind of effect decouples for a sufficiently heavy $M_A$. The DSM
impacts on $m_h$, with the degree determined by several factors,
including $O_{23,32}$. But a large degree never necessarily means
that $O_{21}\tan\beta$ is large (See the left panel of
Fig.~\ref{fig3}). After clarifying these, we employ numerical study
in the rest of this section.

\subsection{Numerical studies}

In the MSSM we use HDECAY~\cite{Djouadi:1997yw} and
CALHEP~\cite{Belyaev:2012qa} to calculae Higgs signatures and stop
decays, respectively. And NMSSMtools 2.3~\cite{NMSSMTools} is used
for the relevant calculations in the NMSSM. In terms of the previous
analysis, we set scanning parameters as the following:
\begin{align}\label{push:R}
{\rm MSSM}:\quad &\tan\beta:[5,\,30],\quad \mu:[100,\,1000]{\rm
\,GeV}, \quad M_A:[300,\,1500]{\rm \,GeV},\cr & m_{\wt q_3}:
[300,\,1000]{\rm \,GeV},\quad m_{\wt u_3}: [800,\,2000]{\rm
\,GeV},\quad A_t:[-3000,\,-1500]{\rm \,GeV}. \cr
 {\rm NMSSM}:\quad
&\tan\beta:[1,\,30],\quad\lambda:[0.1,\,0.72],\quad
\kappa:[0.01,\,0.7], \cr &\mu:[100,\,500]{\rm \,GeV}, \quad
A_{\lambda}:[0,\,3000]{\rm \,GeV},\quad A_{\kappa}:[-600,\,100]{\rm
\,GeV},\cr & m_{\wt q_3},m_{\wt u_3}:[100,\,1000]{\rm \,GeV},\quad
A_t:[-3000,\,0]{\rm \,GeV}.
\end{align}
The SM-like Higgs boson mass is restricted to the region
$123{\rm\,GeV}\lesssim m_h\lesssim 128{\rm\,GeV}$. $\ld\lesssim
0.72$ is required by perturbativity at the GUT scale. In the MSSM
the stop soft masses squared are asymmetric, with $m_{\wt q_3}$
comparatively light so as to keep one stop and sbottom in the lower
mass region. The soft mass squares of the third generation are
relatively small so that the stops and the sbottom can be copiously
produced at the 14 TeV LHC. As for the other sparticles, we fix
their soft masses to be
\begin{align}
&m_{\tilde{b}_R}=3000{\rm \,GeV},\quad m_{\tilde{q}_{1,2}}=2000{\rm
\,GeV},\quad m_{\tilde{l}}=1000 {\rm \,GeV}\cr &M_1=250{\rm
\,GeV},\quad M_1:M_2:M_3=1:2:6.
\end{align}
Thus the light sparticles which are relevant to our study include
stops, the lighter sbottom, Higgsinos and gauginos. Such a setup
keeps the number of parameters as small as possible, and moreover
accords with natural SUSY. Results are displayed in the individual
subsections, including figures from Fig.~\ref{fig1} to
Fig.~\ref{fig6}.

\section{The Stop ensemble at the LHC}

As one of the main object for this article, we will make an anatomy
of the stop system under the condition of a sterile Higgs boson
around 126 GeV. To implement Higgs sterility, we only keep the
points (obtained in the previous section) which satisfy
\begin{align}
0.9\leq R_{\rm VBF}(b\bar b,VV),\,\, R_{\rm gg}(\gamma\gamma,VV)\leq
1.1.
\end{align}
The heavier stop and lighter sbottom, which have not been
extensively studied yet, will gain special attentions here. It is
found that novel signatures from the heavier stop/sbottom cascade
decays may be seen at the LHC. We will focus on the benchmark model
for natural SUSY, the NMSSM, which provides a good laboratory to
study the light stop ensemble facing a sterile Higgs boson. In terms
of the setup for the stop sector, we have the following mass orders:
\begin{align}
m_{\wt t_1}<m_{\wt b_1}\approx m_{\wt Q_3}<m_{\wt t_2}.
\end{align}
Their mass splittings are expected to be large, because a large
$X_t$ is favored by a relatively heavy $m_h$. Of course, altering
the configuration of stop parameters leads to different
distributions of mass spectra and decay widths, but that will not
cause much difference to our discussions on the general features of
the stop ensemble at the LHC.

In the rest of this section, we will first present the distributions of
masses and decays of stops and sbottom, and then explore new signatures
at the LHC. All of the discussions are based on the NMSSM unless otherwise
specified. In fact, even disregarding their intimate connections with
the Higgs boson properties and just for inspecting naturalness alone,
our attempt is meaningful.

\subsection{Decays of two stops and light sbottom}
We now report the distributions of the main decay modes of $\wt t_{1,2}$
and $\wt b_1$, respectively. In the discussion of Higgs
mixing in the NMSSM, we divide it into several distinctive cases.
But decays of stop/sbottom do not show qualitative differences in
different cases, so we only display results of the pulling-down
scenario in this model, which is favored by Higgs sterility.
\begin{description}
  \item[On $\wt t_1$] Distributions of the main decay branching ratios of
  of $\wt t_1$ in Fig.~\ref{fig4}. From it we make a few observations.
  In the lighter stop mass region, $m_{\wt t_1}\lesssim 500$ GeV, the
  mode $\wt t_1\ra b\tilde{\chi}_1^{\pm}$ (via the $\wt t_R$
  component) usually has a lager branching ratio than others,
  such as that of $\wt t_1\ra t \tilde{\chi}^0$. And its LHC bound
  is not strong if the masses of
$\tilde{\chi}^{\pm}_1$ and $\tilde{\chi}^0_1$ are neither
degenerate~\cite{st1} nor hierarchical~\cite{st2}. As a matter of
fact, the current LHC exclusion on light stops is not our concern
here~\cite{Yan}, since that depends on the detailed models, e.g.,
whether $R-$parity is violated or not. In the heavier stop mass
region, $\tilde{t}_1 \to t \tilde{\chi}^0_{i>1}$ has a similar
branching ratio with Br$(\tilde{t}_1 \to b \tilde{\chi}^\pm_{i>1})$,
while other modes are suppressed.

\begin{figure}[htb]
\begin{tabular}{cc}
\includegraphics[width=3.0in]{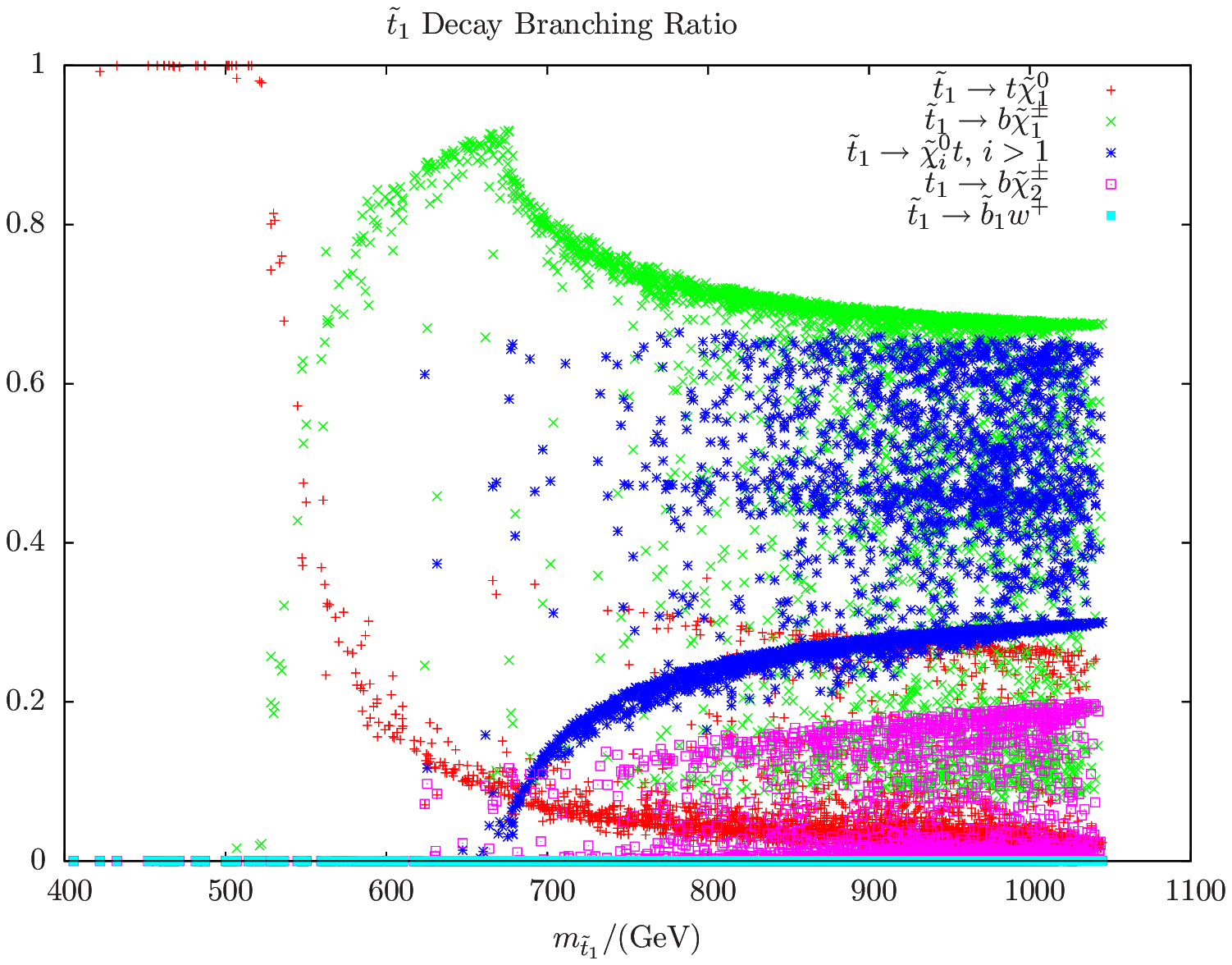}
\includegraphics[width=3.2in]{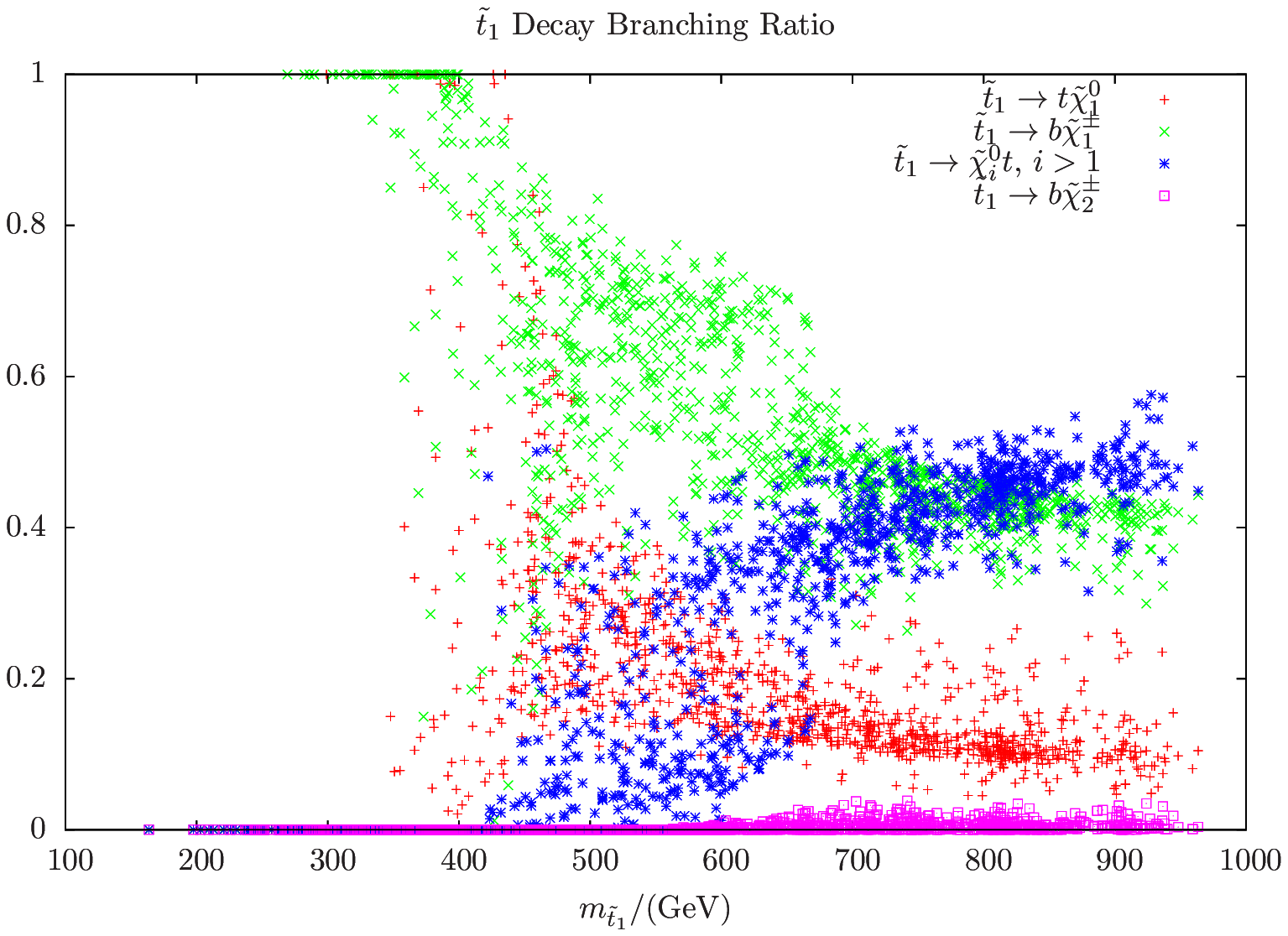}
\end{tabular}
\caption{\label{fig4} Plots of decay branching ratios of $\wt t_1$.
Left panel: MSSM; Right panel: NMSSM in the pulling-down region.
Other scenarios of the NMSSM give similar results, and differ mainly
in the stop mass. So they are not shown explicitly.}
\end{figure}

\item[On $\wt b_1$] In our setup, the sbottom mass can be as low as 200
  GeV. As $\wt t_1$, we keep an open attitude on the LHC bounds on that
  light sbottom. On $\wt b_1$ decays, the modes $\wt b_1\ra\chi^0_1 b$ and $\wt b_1 \ra
 \sum_{i\geq2}\chi^0_ib$ almost take over the lower mass region of
 $\wt b_1$ (below about $\sim$ 400 GeV). While $\wt b_1\ra \chi^\pm_1 t$ and
  $\wt b_1\ra\wt t_1W^\pm$ are dominant over the heavier sbottom
  region. The latter mode is in our interest in the ensuing
  discussions, so we give the analytical expression of its decay width
 at tree level (The complete one-loop correction on it can be found in
 Ref.~\cite{Arhrib:2004tj}):
\begin{align}\label{btw}
\Gamma(\wt b_1 \ra \wt t_1 W)=\f{g_2^2\cos^2\theta_{\wt
t}}{32\pi}\f{m_{\wt b_1 }^3}{m_W^2}\lambda(m_{\wt b_1},m_{\wt
t_1},m_W)^{3/2}.
\end{align}
with
\begin{align}
\ld(x,y,z)\equiv\left[ 1-\L \f{y+z}{x}\R^2\right]\left[  1-\L
\f{y-z}{x}\R^2\right].
\end{align}
So the relative weights of these two modes are sensitive to the
constituent of $\wt t_1$ and the mass
 splitting between $\wt t_1$ and ${\wt b_1}$. As $\wt b_1$ becomes
 sufficiently heavy (typically heavier than 700 GeV for our choice of
 wino mass, 500 GeV), its decays to $\chi_2^\pm t$ has a branching
 ratio a few tens of percents.

\begin{figure}[htb]
\begin{tabular}{cc}
\includegraphics[width=3.0in]{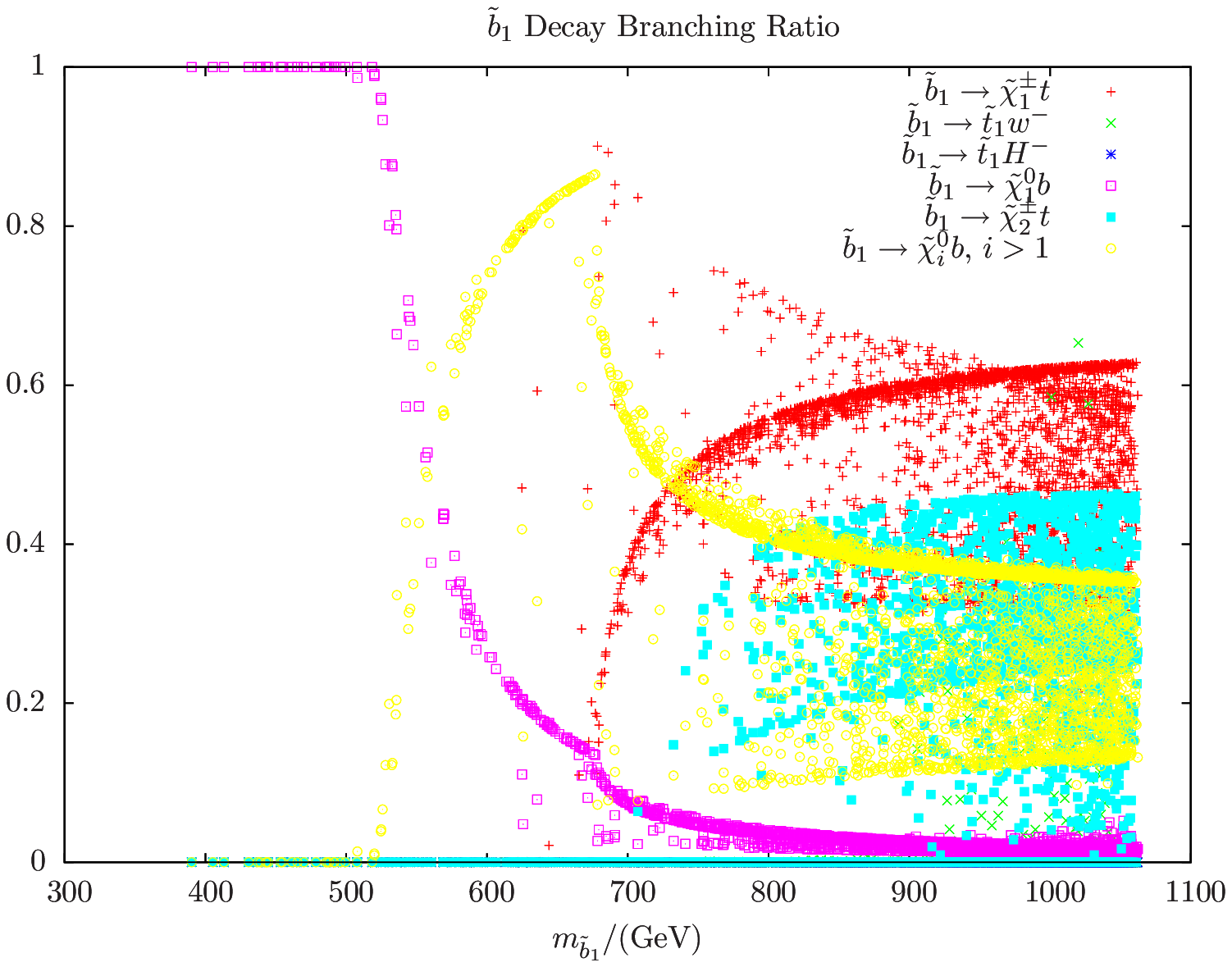}
\includegraphics[width=3.2in]{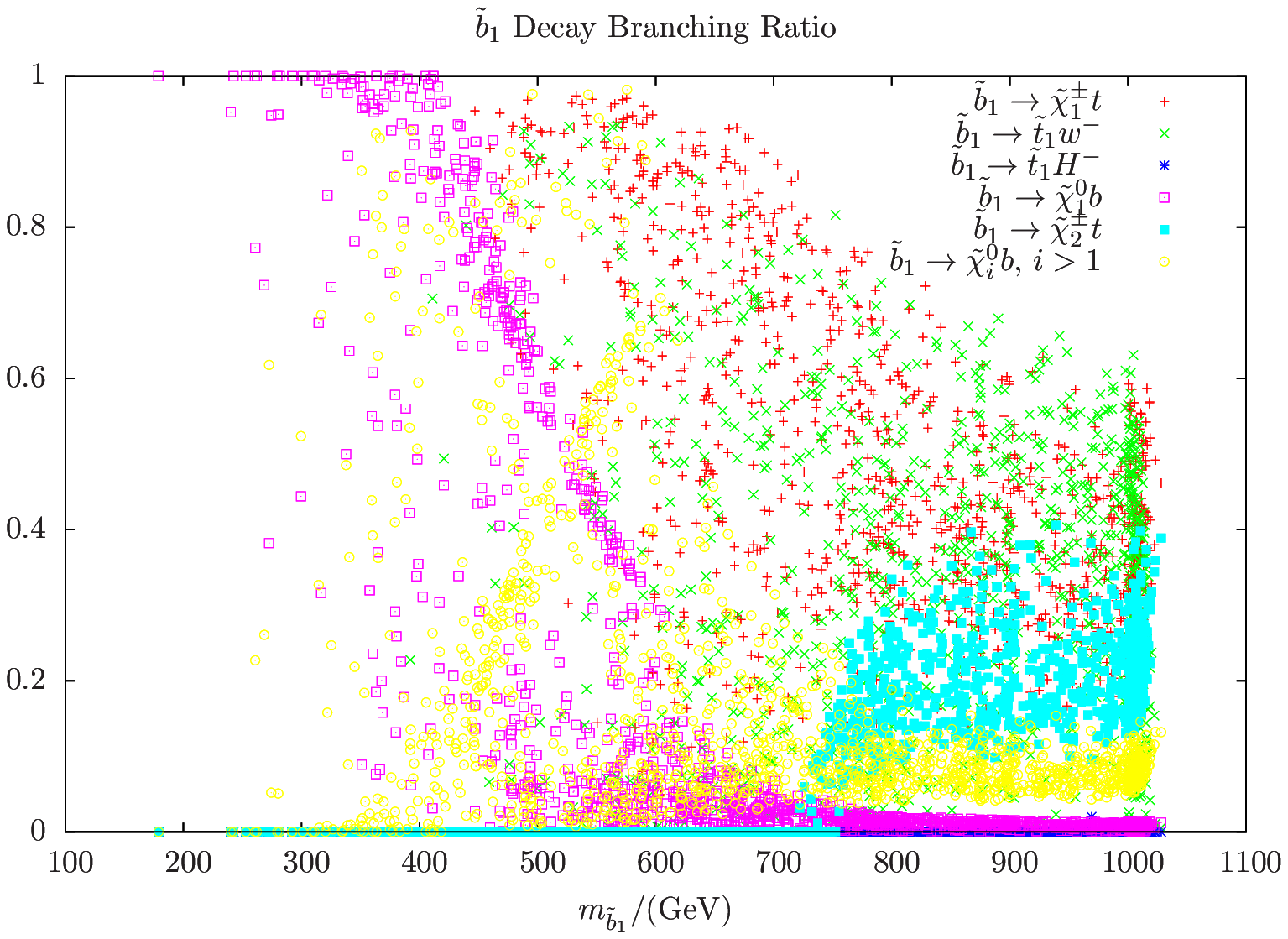}
\end{tabular}
\caption{\label{fig5} Plots of decay branching ratios of $\wt b_1$.
Left: MSSM; Right: NMSSM.}
\end{figure}

\item[On $\wt t_2$] It is the heaviest particle (with mass roughly
above 600 GeV) of the stop ensemble, and consequently it possesses a
rich decay table. That may impede the discovery of this particle due
to the suppressed decay branching ratios of the individual channels.
From Fig.~\ref{fig6} we see that, the conventional decay modes,
i.e., to neutralinos and charginos, usually are subdominant
(typically with branching ratios less than $20\%$), except that $\wt
t_2\ra \sum_{i\geq2}\chi^0_it$ takes up a larger branching ratio.
Remarkably, the interesting modes $\wt t_1Z/h$ and $\wt b_1W^\pm$
have substantial branching ratios. For illustration, the partial
decay widths of $\wt t_2$ to $\wt t_1$ plus $Z$ and $h$ are
respectively given by
\begin{align}\label{}
\Gamma(\wt t_2\ra \wt
t_1Z/h)&\approx\f{g_2^2}{\cos^2\theta_W}\f{\sin^22\theta_{\wt
t}}{256\pi}\f{m_{\wt t_2}^3}{m_Z^2}\ld^{3/2}(m_{\wt t_2},m_{\wt
t_1},m_Z),\\
&\approx\f{\cos^22\theta_{\wt t}}{16\pi}\L\f{y_t^2A_t^2}{m_{\wt
t}^2} \f{m_{\wt t_1}}{m_{\wt t_2}}\R m_{\wt t_2}.
\end{align}
where we have taken $H_u^0\sim h$. The $Z-$mode favors a large
left-right (LR) stop mixing while the $h-$mode, which mainly is
induced by the trilinear soft term $(y_tA_t \wt Q_3H_u\wt
U_3^c+c.c.)$, favors LR stops decoupling, says due to hierarchal
stop soft masses squared. From Fig.~\ref{fig6} we find that, Br$(\wt
t_2\ra \wt t_1Z)\sim 30\%$ in the total mass region of $\wt t_2$,
and Br$(\wt t_2\ra \wt t_1h)$ almost evenly scatters below the
$30\%$ line. As for $\Gamma(\wt t_2\ra \wt b_1W)$, it can be
obtained in analogous to Eq.~(\ref{btw}) after the replacements
$\cos\theta_{\wt t}\ra \sin\theta_{\wt t}$ and $\wt b_1\ra \wt t_2$,
$\wt t_1\ra \wt b_1$. And its branching ratio is smaller than
$40\%$.

\begin{figure}[htb]
\begin{tabular}{cc}
\includegraphics[width=3.0in]{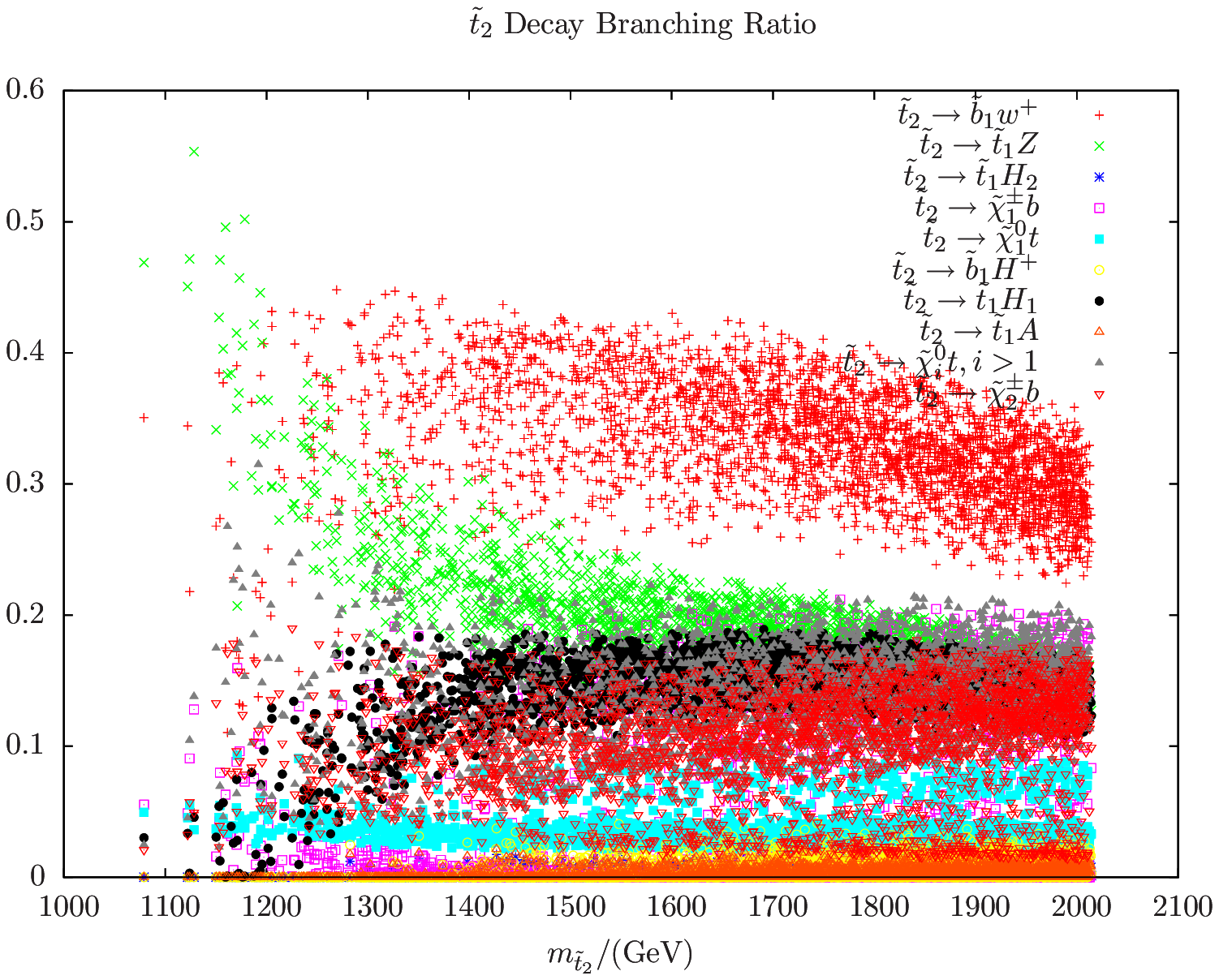}
\includegraphics[width=3.2in]{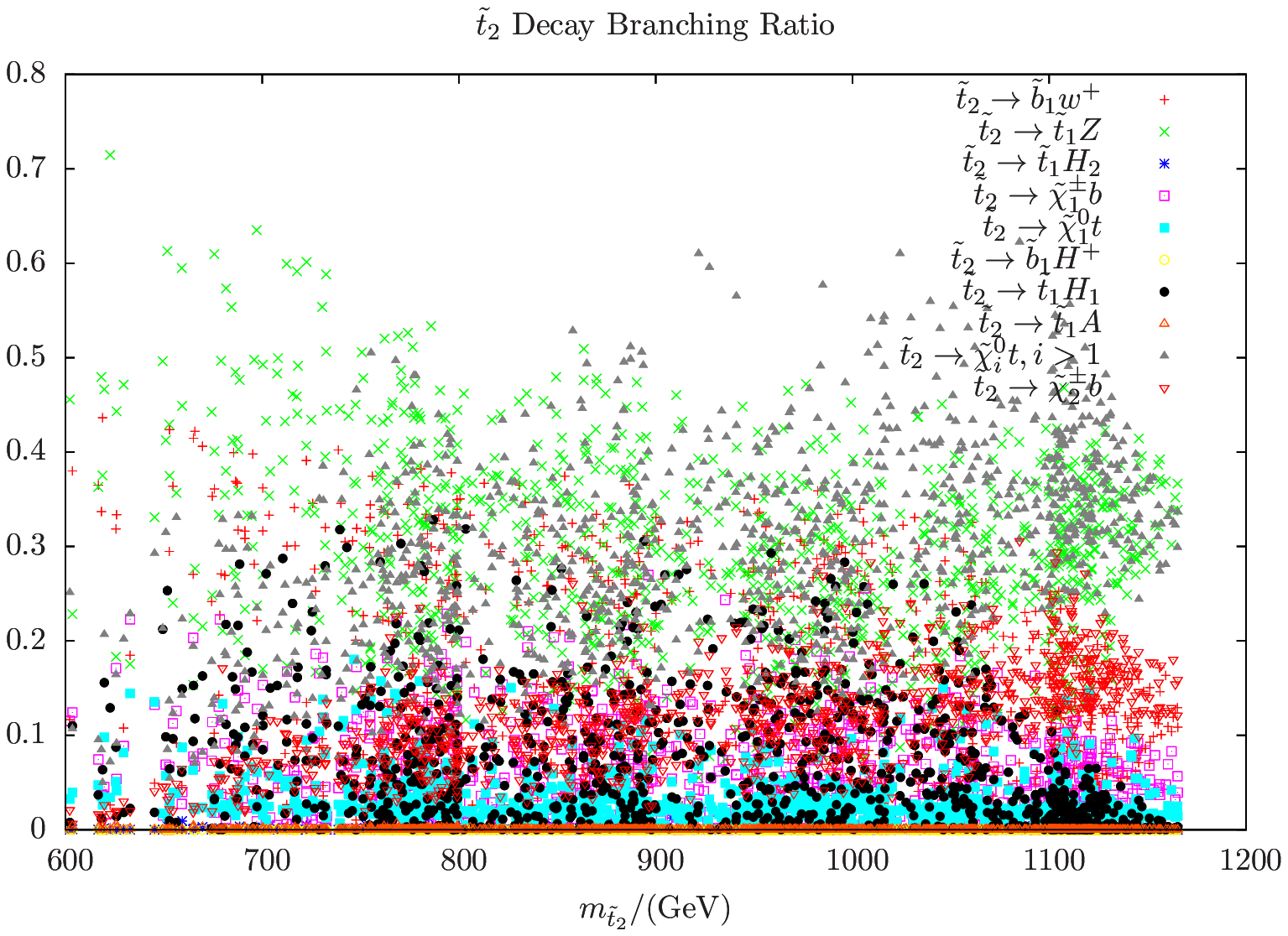}
\end{tabular}
\caption{\label{fig6} Plots of decay branching ratios of $\wt t_2$.
Left: MSSM; Right: NMSSM.}
\end{figure}
\end{description}

\subsection{Explore the heavier stop and sbottom LHC signatures}

With the aid of the results in the previous subsection, we now
attempt to preliminarily explore the characteristic signatures for
the stop ensemble at the LHC. We will not devote ourself to $\wt
t_1$, which has been the focus of many works. The decays of heavier
states $\wt t_2$ and $\wt b_1$ may give rise to novel collider
signatures, which potentially provide a way to probe the stop
ensemble rather than $\wt t_1$ alone. Signatures of stops/sbottom
strongly depend on the decay chains of neutralinos/charginos, which
however are not specified in this work. They can be very different
in different SUSY scenarios. For example, in certain
$R-$parity-violating SUSY, the large missing energy is absent and
consequently most of the current stop searches are invalid. In what
follows we present several categories of signatures.
\begin{description}
    \item[Same-sign dilepton (SSDL) \& Multi-leptons (MLs)]
    Signatures containing SSDL or MLs are common
to several channels, thanks to the hard $W$ or/and $Z$ bosons
generated during the cascade decays of the heavier stop/sbottom to
the lighter states. SSDL is rare in the SM, so it provides a
promising avenue for observing the additional third family colored
sparticles.

Considering the relatively heavy $\wt b_1$ pair production and at
least one $\wt b_1$ decays along the chain (We use superscript
``$\pm"$ to denote the sign of charge, discarding its value):
\begin{align}\label{sbl}
\wt b_1^-&\ra \wt t_1(\ra \sum_{i\geq1}\tilde{\chi}^0_i+W^++b^-)+W^-,
\end{align}
which produces a pair of opposite-sign dibosons. According to the
previous numerical results, the other sbottom $\wt b_1^+$ dominantly
decays into either $\wt t_1^-W^+$ or $\wt\chi^+ t^-$. Combining with
the products of $\wt b_1^-$ decay, in any case one gets the
same-sign dibosons with an appreciable cross section. Actually, we
can even get $W^+W^+$ plus $W^-W^-$, but with a significantly
reduced cross section. $\wt t_2$ decay is also a rich source of
SSDL. Similarly, considering the pair production of $\wt t_2$,
followed by at least one of them decays as:
\begin{align}\label{st2}
\wt t_2^+&\ra \wt t_1(\ra \tilde{\chi}^0+W^++b^-)+Z,\quad \cr &\ra \wt b_1\left[\wt t_1(\ra \tilde{\chi}^0+W^++b^-)+W^-\right]+W^+.
\end{align}
Each chain itself produces SSDL, and thus if we inclusively observe
the SSDL, the LHC sensitivity can be substantially improved.

We would like to give several comments. In the first, the
$W/Z-$richness in the above decay chains means that final states may
be lepton rich, so multi-leptons (MLS) deserve attentions. Next, we
do not take the neutralinos and charginos decays into account.
Actually, charged leptons are likely to be produced, mediated by the
on- or off-shell $W(Z)-$bosons, in the $\chi^\pm(\chi_i^0)$ cascade
decays. So $\wt t_2\ra \sum_{i\geq2}\chi^0t$ and $\wt b_1\ra
\chi_1^\pm t$, which have large branching ratios, provide SSDL also.
Finally, the current CMS searches for the SSDL accompanied by at
least two $b-$jets~\cite{Chatrchyan:2012paa}, and signatures are
divided into categories both with and without large MET. SSDL from
$\wt t_2/\wt b_1$ decay satisfies the criterion and is thus subject
to the CMS constraint. In some case, the $\sqrt{s}=8$ TeV and the
$\mathcal{L}=$ 10.5 fb$^{-1}$ data has already set a lower bound of
450 GeV on $\wt b_1$~\cite{Chatrchyan:2012paa}.
\begin{figure}[htb]
\includegraphics[width=3.2in]{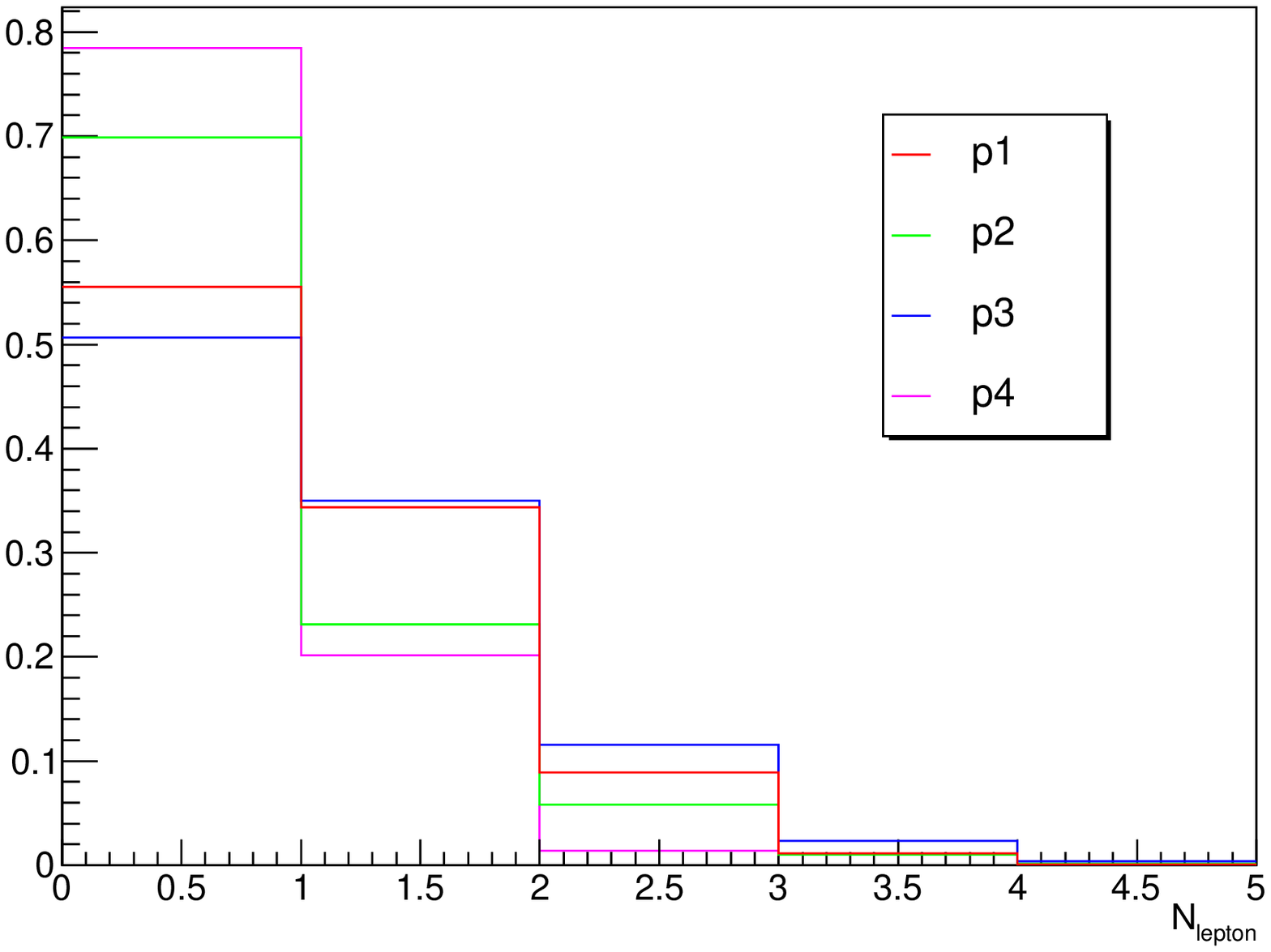}
\includegraphics[width=3.2in]{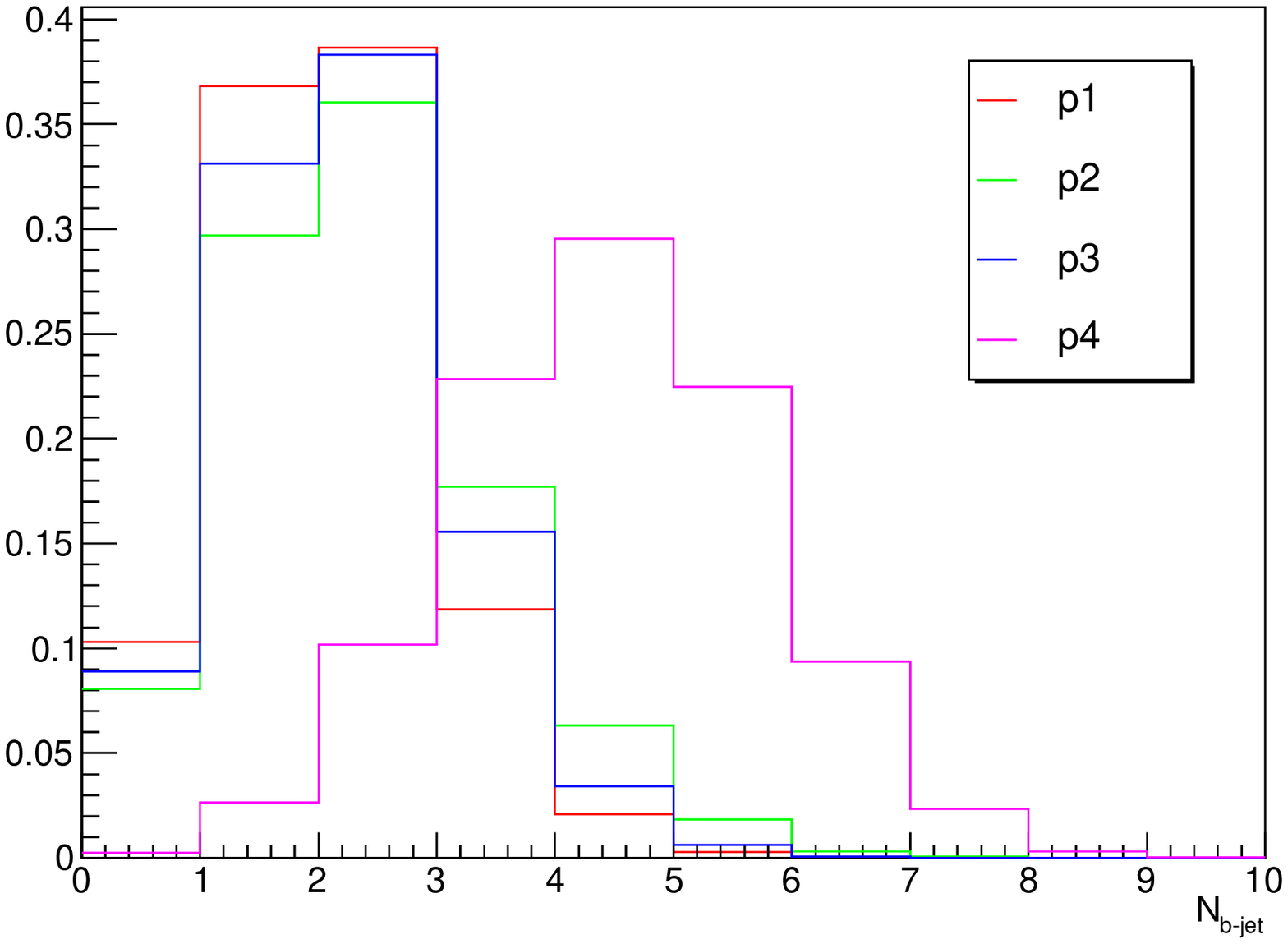}
\caption{\label{nl3} Distribution of the numbers of leptons
$N_{lepton}$ and $b-$jets $N_{b-jet}$ for the four benchmark points.
The vertical axis denotes the number of events, in unit 50000 (same
in Fig.~\ref{pttop}).}
\end{figure}

\item[Multi $b-$jets] Top quark and $Z/h$ are sources of $b-$jet. So, it is expected that multi $b-$jets (no less than 3) signature
is produced in the stop ensemble. This signature alone is powerful.
For example, it helps to discover $t'$ with mass  $\lesssim$ 550 GeV
at 5$\sigma$ level~\cite{Harigaya:2012ir}. Here, it can be further
strengthened by assistant cuts such as a large MET and thus
vigorously probes the heavier stop/sbottom. As before, we do not
need to specify the neutralino/chagino decays.

This signature is especially suited for searching $\wt t_2$. Still
considering the $\wt t_2$ pair production, the pattern of subsequent
decay is
\begin{align}\label{}
\wt t_2\ra \wt t_{1}+Z/h\ra 3b+X~ {\rm~and}~~ \wt t_2\ra b+X,
\end{align}
Since $\wt t_2$ decay produces at least one hard $b-$jet, so Br$(\wt
t_2\ra b+X)$ does not suffer suppression from branching ratios.
Similar search strategy has been adopted in
Ref.~\cite{Berenstein:2012fc,Ghosh:2013qga}, where the jet
substructure of $b\bar b$ from $h$ or $Z$ decay is used to enhance
the signal sensitivity. Pair production of $\wt b_1$ can not give
rise to the multi hard $b-$jets signature except for taking into
account the $Z/h$ bosons from the heavier neutralino decays.

We note that the signature 2$b-$jets+MET has been utilized by CMS
and ATLAS~\cite{2b:MET} to search sbottom with decay mode $\wt
b_1\ra \chi_1^0b$. Although it is a strong signature of $\wt t_2/\wt
b_1$, the present searches hardly constrain the stop ensemble in
this paper. The reason is that, on the one hand, the mode $\wt b_1\ra
\chi_1^0 b$ is subdominant for heavier $\wt b_1$; On the other hand,
to suppress the huge $t\bar t$ background, they vetoe leptons which
however are generic from the $\wt t_2/\wt b_1$ decays.

\item[Boosted tops] Top quarks appear in the most decay chains of
$\wt t_2/\wt b_1$. Thereby, for the stop ensemble lies in the
heavier region, says close to the TeV scale,  signatures containing
boosted tops are well expected. Boosted tops can be produced from
the primary of $\wt t_2/\wt b_1$, via $\wt t_2\ra t\wt\chi_i^0$ and
$\wt b_1\ra \wt\chi_i^\pm b$, or from their secondary decay as shown
in the benchmark points. But the latter case only produces
moderately boosted tops with $p_T\sim200$ GeV, given $\wt t_1$
around 500 GeV. They can be tagged using
heptoptagger~\cite{Plehn:2010st}. For $p_T \gtrsim 200$ GeV, the top
tagger efficiency is around 30\% or even higher~\cite{Plehn:2010st}.
However, top-tagging alone fails to kill the huge backgrounds from
$t\bar{t}$ production. So we may need the help from other variables,
e.g., $m_{T2}$. Because of the heaviness of mother particles, the
signatures have much larger $m_{T2}$ than that of the $t\bar{t}$
background~\cite{Chakraborty:2013moa}.

\begin{figure}[htb]
\begin{tabular}{cc}
\includegraphics[width=4.2in]{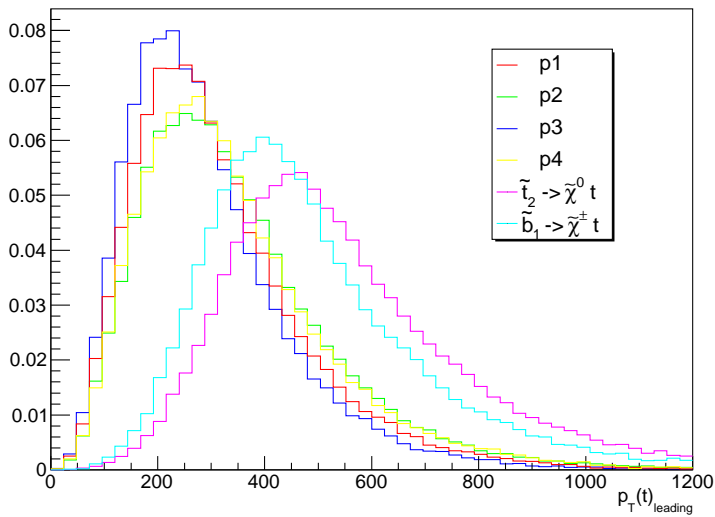}
\end{tabular}
\caption{\label{pttop} Distribution of $p_T$ of top quark in each
benchmark points and from the ordinary decay channels, i.e., these
with primary top quark. Here both $\wt\chi^0$ and $\wt \chi^\pm$
have mass 100 GeV. }
\end{figure}

\end{description}

To form an initial impression on the LHC prospects of the
characterized signatures originating from decays between stop and
sbottom, we consider four benchmark points, which are listed in the
second and third columns of Table.~\ref{bp3}. Each step along the
decay chain has been assumed to have a 100$\%$ branching ratio,
except for the well known particles $t$, $W$ and $Z$, which decay in
PYTHIA.
For each point, 50000 events at the 14 TeV LHC are generated by
MadGraph5~\cite{Alwall:2011uj}, and passed to
PYTHIA6~\cite{Sjostrand:2006za} for particle decay and parton
shower. The detector effects are implemented by
Delphes3~\cite{deFavereau:2013fsa}.

\begin{table}
    \begin{tabular}{|r|c|c|c|c|}
    \hline
    & Channel & Masses & $R_{\rm SSDL}$ & $N_{\rm SSDL}/100fb^{-1}$ \\ \hline
    $p_1$ & $\tilde{b}_1 \to \tilde{t}_1 W^- \to (t \tilde{\chi}^0_1) W^-$ &  $m_{\tilde{b}_1}=800$ GeV & $\frac{1680}{50000}$ & 116.9 \\ \hline
    $p_2$ & $\tilde{t}_2 \to \tilde{t}_1 Z \to (t \tilde{\chi}^0_1) Z$ & $m_{\tilde{t}_2}=900$ GeV & $\frac{477}{50000}$ & 15.0 \\ \hline
    $p_3$& $\tilde{t}_2 \to \tilde{b}_1 W^+ \to (\tilde{t}_1 W^-) W^+ \to (t \tilde{\chi}^0_1 W^-) W^+$ &  $m_{\tilde{t}_2}=900$ GeV, \ $m_{\tilde{b}_1}=700$ GeV & $\frac{2817}{50000}$ &88.5 \\ \hline
    $p_4$& $\tilde{t}_2 \to \tilde{t}_1 h \to (t \tilde{\chi}^0_1) (b \bar{b})$ & $m_{\tilde{t}_2}=900$ GeV & $\frac{4}{50000}$ & 0.1 \\ \hline
    \end{tabular}
    \caption{\label{bp3} $m_{\chi^0_1}=100$ GeV, $m_{\tilde{t}_1}=400$ GeV}
\end{table}

We start from SSDL. We adopt the ATLAS definition of SSDL~\cite{atlas2ssl},
which requires two leading isolated leptons with $p_T>20$ GeV and
$|\eta|<2.47$ for electron while $|\eta|<2.4$ for muon which carries
the same electric charge with the electron. Lepton isolation
requires that, inside a cone of $R = 0.15$ around this lepton, the
scalar sum of $p_T$ of the final partilces is less than 10\% of
$p_{T,lepton}$. The rates of SSDL in each benchmark point are given
in Table~\ref{bp3}, the fourth column. We can understand the results
via the naive estimation like
\begin{align}
R_{\rm SSDL}(p_1)\simeq2{\rm Br}({W_\ell})^2{\cal P}_1,
\end{align}
with the $W_\ell$ and $Z_{\ell}$ leptonic decay branching ratios
about $1/5$ and $1/10$, respectively. Then it is seen that the
probability of SSDL ${\cal P}_1\sim50\%$, a remarkably high
probability. Given SSDL rates, we estimate the corresponding numbers
of events at the 14 TeV LHC with integrated luminosity 100 fb$^{-1}$
(We calculate the production cross sections
using~\cite{Beenakker:1996ed}). The results are listed in the last
column of Table~\ref{bp3}. As one can see, $p_1$ and $p_3$, namely
both $\wt b_1$ and $\wt t_2$, have a good chance to be discovered.
As for the MLs, its rate is suppressed by the decay branching ratios
and thus is not that attractive, see the right panel of
Fig.~\ref{nl3}. We now turn our attention to the multi $b-$jets. We
include a $b$-tagging efficiency of 70\% and a probability of 10\%
and $1\%$ for mis-tagging a charm quark and other light quarks,
respectively. The distributions of $b-$jets numbers $N_b$ are
displayed in the left panel of Fig.~\ref{nl3}. From it one can see
that, all the benchmark points are $b-$rich ($N_b \geq 2$), and
especially, the number of $b$-jets of $p_4$ peaks at 4. Finally, we
plot the $p_T$ distribution ot top quark, in Fig.~\ref{pttop}. It
shows that, as expected, top from secondary decay is moderately
boosted, with (leading top) $p_T$ slightly above $m_{\wt t_1}$/2,
while the primary top quark is highly boosted with $p_T$ peaks at
half of the mother particle mass. In summary, the stop ensemble
closing 1 TeV can be probed via SSDL, multi $b-$jets or boosted top.
But here we only make the preliminary analysis of the signature
properties, and the quantitative collider study, like improved cuts
and backgrounds analysis, is left for future work.

\section{Conclusions and discussions}

As the LHC data accumulates, it is likely to show us a sterile Higgs
boson. That is to say, its (main) signature strengths deviating from
the SM predictions are within the experimental resolution
($\lesssim10\%$). Recalling that in the SSM Higgs couplings are
often modified by mixing and stops, Higgs sterility should have a
deep implication on the Higgs and stop sector. We analyzed that based
on two benchmark models:
\begin{itemize}
  \item  In the nearly decoupling region of MSSM, the doublet-doublet
  mixing effect is universal up to an individual enhancement in
  $C_{hb\bar b}$, by $2m_Z^2/M_A^2$. Higgs sterility then places a bound:
  $M_A \gtrsim 900$ GeV. Since $m_h\simeq126$ GeV relies
on a heavy stop sector, then to get a relatively light stop we
should turn to large stop mixing or/and asymmetric stop soft mass
squared. Such cases are subject to constraint from the Higgs
sterility. But generically sterility does not mean much to the stop
sector.
  \item Similarly, in the NMSSM violation of universality of the
  Higgs mixing effect is encoded in $C_{hb\bar b}$. However,
  here $C_{hb\bar b}$ can be made either smaller or larger than 1,
due to the distortion of doublet-doublet
  mixing effect by DSM. Interestingly, given a large $\tan\beta$
  the DSM effect may not simply vanish as $M_A$ increases.
  Moreover, depending on the structure of the Higgs sector,
  the DSM effect can push-up $m_h$ or pull-down $m_h$. In the
former scenario, the amount of pushing-up is less than $\sim$ 5 GeV
due to sterility. In particular, we revise to the pushing-up region
with a large $\tan\beta$ and moderately small $\ld$, which may help
to embed the low energy NMSSM into the (semi)constrained
form~\cite{CNMSSM}. In the pulling-down scenario, Higgs sterility is
automatically implemented, because to weaken the pulling-down effect
DSM is strongly favored to be small. In any case, stops in the NMSSM
are allowed to be comparatively light, so Higgs sterility both
directly and indirectly constrains them.
\end{itemize}
We have to emphasize that we here focus on the tree-level analysis.
The full supersymmetric QCD correction (In this paper it is only
partially included because we fixed many relevant parameters, like
the gluino mass.) may change $C_{hb\bar b}$
substantially~\cite{Liu:2012qu,Cahill-Rowley:2013vfa}. However,
radiative correction strongly depends on the total soft spectrum,
which renders a generic prediction very difficult.

We also studied the LHC features of the whole stop sector, rather than
merely the lightest stop (A work in this inspirit has
appeared~\cite{Kribs:2013lua}.), allowed by a 126 GeV sterile Higgs
boson. We first made a detailed numerical analysis of the stop
sector of the NMSSM, including the mass and decay distributions of
$\wt t_2$ and $\wt b_1$. Then we propose several promising
signatures for discovering the heavier stop and sbottom. Due to the
cascade decays among stops and sbottom, same sign leptons and
multi-$b$ jets are characterized signatures and have promising
prospect at the future LHC.

To end up this work, we add several remarks. First, although a lot
of papers have studied the mixing effect(s) in the (N)MSSM, our
paper reveals their most remarkable features and clarifies some
points which seem to be unclear in the literatures. Additionally,
the idea of using Higgs sterility to constrain new physics, of
course, can be generalized to many other contexts where Higgs
couplings are
modified~\cite{Batell:2012mj,Chpoi:2013wga,Lopez-Val:2013yba}. As
the final remark, we would like to stress that our discussions on
characteristic signatures of the heavier stop/light sbottom actually
are based on SUSY with less fine-tuning, so our work may open a new
window to probe natural SUSY. But the results presented in this
paper are preliminary, and their actual prospects need more detailed
LHC analysis, and we leave it for an open question.

\section{Acknowledgements}

We would like to thank Taoli Cheng for helpful discussions. This
research was supported in part by the China Postdoctoral Science
Foundation under grant numbers 2013M530006 (KZ), and by the Natural Science
Foundation of China under grant numbers 10821504, 11075194,
11135003, and 11275246.

\appendix
\section{Higgs effective couplings}\label{convention}

In this appendix we briefly introduce how to construct effective
couplings of the SM-like Higgs boson $h_{\rm SM}$. We start from the
Lagrangian with tree-level couplings only:
 \begin{align}\label{tree}
{\cal L}_{\rm tree}\supset &
r_{i,Z}\f{M_Z^2}{\sqrt{2}v}H_iZZ+r_{i,W}\f{\sqrt{2}M_W^2}{v}H_iW^+W^-
-r_{i,f}\f{m_f}{\sqrt{2} v}H_i\bar ff-r_{i,S}\f{\sqrt{2}m_S^2}{
v}H_i S^\dagger S,
\end{align}
with $v\approx174$ GeV. In the NMSSM, we have $h_{\rm SM}=H_i$ with
$i=1$ or 2. Here $f$ and $S$ denote a Dirac fermion and complex
scalar, respectively. For the particles belonging to the SM, the
dimensionless parameters $r_{i,V}$, etc., can measure the deviations
of $H_i$ from $h_{\rm SM}$. They are supposed to slide to 1 when
$H_i$ exactly coincides with $h_{\rm SM}$.

In Eq.~(\ref{tree}), particles carrying QCD or/and QED charges
generate Higgs effective couplings to gluons and photons at loop
level. They are crucial to the detection of Higgs boson at the LHC,
and incorporated through the following dimension-five
operators~\cite{shifman}:
 \begin{align}\label{O:5}
{\cal L}_{\rm loop}=r_{i,g}\f{\alpha_s}{12\sqrt{2}
v}H_iG_{\mu\nu}^aG^{a,\mu\nu}+r_{i,\gamma}\f{\alpha}{\sqrt{2}
v}H_iF_{\mu\nu}F^{\mu\nu}.
\end{align}
Note that in this notation $r_{i,g}$ and $r_{i,\gamma}$ are not 1 in
the SM limit. The operator coefficients can be calculated in terms
of the following formulas (See Ref.~\cite{Carmi:2012in} and
references therein):
\begin{align}
r_g=&\f{C_2(r_s)}{2}r_s {\cal A}_s(\tau_s)+2C_2(r_f)r_f{\cal
A}_f(\tau_f),\cr r_\gamma=&\f{N(r_s)Q_s^2}{24}r_s {\cal
A}_s(\tau_s)+\f{N(r_f)Q_f^2}{6}r_f {\cal
A}_f(\tau_f)-\f{7Q_V^2}{8}r_V{\cal A}_V(\tau_V),
\end{align}
where $C_2(r)$ and $N(r)$ are the quadratic Casimir and number of
colors of the representation $r$ under $SU(3)_C$. For a heavy
particle with $\tau\equiv m_h^2/4m^2\ll 1$, its loop function ${\cal
A}\ra 1$ and the corresponding contribution is then fixed up to the
parameter $r$. Within the SM, the top quark and $W-$boson dominantly
account for Eq.~(\ref{O:5}). In the (N)MSSM we have
 \begin{align}\label{rggamma}
r_{i,g}\approx&1.03\,r_{i,t}-0.06\,r_{i,b}+\delta r_{i,g}(\rm
stops),\cr
 r_{i,\gamma}\approx &\f{2}{9}\times1.03\,r_{i,t}-1.04\,r_{i,V}+
 \delta r_{i,\gamma}(\rm
stops,~chargino).
\end{align}
To get them we have taken $m_{H_i}\simeq$126 GeV. For the exact SM
Higgs boson, i.e., $H_i=h_{\rm SM}$ we have $r_{{\rm SM},g}=0.97$
and $r_{{\rm SM},\gamma}=-0.81$.

To compare with experimental data, it is convenient to express Higgs
signature strengths in terms of $r$. For example, for
$X=(2\gamma,~VV,~b\bar b,...)$ from the gluon fusion channel we have
\begin{align}
R_{gg}^{H_i}(X)\equiv \f{\Gamma(H_i\ra gg){\rm Br}(H_i\ra
X)}{\Gamma(h_{\rm SM}\ra gg){\rm Br}(h_{\rm SM}\ra X)}
=\f{r_{i,g}^2}{r_{{\rm SM},g}^2}\f{r_{i,X}^2}{r_{{\rm SM},X}^2}
\f{1}{B_{\rm tot}},
\end{align}
with $C_{\rm tot}$ the ratio of total decay widths, i.e.,
${\Gamma_{H_i}}/{\Gamma_{h_{\rm SM}}}$. Signature strengths from
other channels can be defined similarly. In literatures such as the
NMSSMTools~\cite{NMSSMTools}, the reduced couplings $C_{i,X}\equiv
r_{i,X}/r_{{\rm SM},X}$ are used. With this notation,
$R_{gg}^{H_i}(X)=C_{i,g}^2C_{i,X}^2/B_{\rm tot}$ with
\begin{align}
B_{\rm tot}\approx 0.64
C_{i,b}^2+0.24C_{i,V}^2+0.09C_{i,g}^2+0.03C_{i,t}^2\leq1.
\end{align}
To derive it we have used: ${\rm Br}(h_{\rm SM}\ra b\bar b+\tau\bar
\tau)=0.64$, ${\rm Br}(h_{\rm SM}\ra WW^*+ZZ^*)=0.24$, ${\rm
Br}(h_{\rm SM}\ra gg)=0.085$ and ${\rm Br}(h_{\rm SM}\ra c\bar
c)=0.027$.

\section{The Higgs and stop mass square matrices }\label{MH2}

In the basis $(S_1,S_2,S_3)$ defined in the text, the elements of
the CP-even Higgs mass square matrix $M_S^2$ are given by
\begin{align}\label{Higgs:even}
&(M_S^2)_{11}=M_A^2+(m_Z^2-\ld^2v^2)\sin^22\beta,\quad\quad
(M_S^2)_{12}=-\f{1}{2}(m_Z^2-\ld^2v^2)\sin4\beta,\cr
&(M_S^2)_{13}=-\f{1}{2}(M_A^2\sin2\beta+2\ld\kappa
v_s^2)\cos2\beta\f{v}{v_s},\quad\quad
(M_S^2)_{22}=m_Z^2\cos^22\beta+\ld^2v^2\sin^22\beta,\cr
&(M_S^2)_{23}=\f{1}{2}(4\ld^2v_s^2-M_A^2\sin^22\beta-2\ld\kappa
v_s^2\sin2\beta)\f{v}{v_s},\cr
&(M_S^2)_{33}=\f{1}{4}M_A^2\sin^22\beta\L\f{v}{v_s}\R^2+4\kappa^2v_s^2+\kappa
A_\kappa v_s-\f{1}{2}\ld\kappa v^2\sin2\beta,
\end{align}
where  $M_A^2=2\ld v_s(A_\ld+\kappa v_s)/\sin2\beta$. Using it, we
can rewrite $(M_S^2)_{23,33}$ as
\begin{align}\label{Higgs:even1}
 &(M_S^2)_{23}=2\ld
\mu\nu\left[1-\L\f{A_\ld}{2\mu}+\f{\kappa}{\ld}\R\sin2\beta\right],\cr
&(M_S^2)_{33}=\ld^2v^2\f{A_\ld}{2\mu}\sin2\beta+4\f{\kappa^2}{\ld^2}\mu^2+\f{\kappa}{\ld}A_\kappa
\mu.
\end{align}

The stop sector has three parameters, casted in the stop mass square
matrix ${M_S^2}_{stop}$. In the basis $(\wt t_R,\,\wt t_L)$, it
takes the form of
\begin{align}\label{stop}
{M_S^2}_{stop}=\left(\begin{array}{cc}
            m_{\wt t_R}^2+m_t^2-(v_u^2-v_d^2)g_1^2/3& m_t(A_t-\mu\cot\beta)\\
            & m_{\wt t_L}^2+m_t^2+(v_u^2-v_d^2)\L g_1^2/12+g_2^2/4\R
                        \end{array}\right).
\end{align}
We define the first and second diagonal entries of ${\cal
M}^2_{stop}$ as $m_{RR}^2$ and $m_{LL}^2$, respectively. The mass
eigenstates are denoted as $\wt t_{1,2}$, and the corresponding
eigenvalues are
\begin{align}\label{}
m_{\wt t_{1,2}}^2=\f{1}{2}\left[\L m_{LL}^2+m_{RR}^2\R\mp \sqrt{\L
m_{LL}^2-m_{RR}^2\R^2+4X_t^2m_t^2}\right],
\end{align}
with $X_t\equiv A_t-\mu\tan\beta$. The flavor and mass eigenstates
are related by
\begin{align}\label{}
\wt t_{L}=\cos{\theta_{\wt t}}\wt t_{1}-\sin\theta_{\wt t}\wt
t_2,\quad \wt t_{R}=\sin{\theta_{\wt t}}\wt t_{1}+\cos\theta_{\wt
t}\wt t_2,
\end{align}
with the stop mixing angle $\theta_{\wt t}$ defined through
$\tan2\theta_{\wt t}=2X_t m_t/(m_{LL}^2-m^2_{RR})$. Thereby, the
degeneracy between $m_{RR}^2$ and $m_{LL}^2$, or/and large
left-right stop mixing $X_t$ lead to $\theta_{\wt t}\ra \pi/4$.


\vspace{-.3cm}
\begin{thebibliography}{99}



\bibitem{LHC:Higgs}
G. Aad et al. [ATLAS Collaboration], Phys. Lett. B 716, 1 (2012); S.
Chatrchyan et al. [CMS Collaboration], Phys. Lett. B 716, 30 (2012).

\bibitem{Mohr:2013eba}
  N.~Mohr [on behalf of the CMS Collaboration],
  arXiv:1307.5745 [hep-ex].

\bibitem{Peskin:2012we}
  M.~E.~Peskin,
  arXiv:1207.2516.

\bibitem{Ellwanger:2009dp}
  U.~Ellwanger, C.~Hugonie and A.~M.~Teixeira,
  Phys.\ Rept.\  {\bf 496}, 1 (2010).

\bibitem{Kang:2013wm}
  Z.~Kang, Y.~Liu and G.~-Z.~Ning,
  arXiv:1301.2204.


\bibitem{Basak:2012bd}
 J.~R.~Espinosa and M.~Quiros,
  Phys.\ Lett.\ B {\bf 279}, 92 (1992);
  T.~Basak and S.~Mohanty,
  Phys.\ Rev.\ D {\bf 86}, 075031 (2012);
  A.~Delgado, G.~Nardini and M.~Quiros,
  arXiv:1207.6596.

\bibitem{Djouadi:2005gj}
  A.~Djouadi,
  Phys.\ Rept.\  {\bf 459}, 1 (2008).

\bibitem{Arbey:2013jla}
  A.~Arbey, M.~Battaglia and F.~Mahmoudi,
  Phys.\ Rev.\ D {\bf 88}, 015007 (2013).

\bibitem{Carmi:2012in}
  D.~Carmi, A.~Falkowski, E.~Kuflik, T.~Volansky and J.~Zupan,
  JHEP {\bf 1210}, 196 (2012).

\bibitem{Ajaib:2012eb}
  M.~A.~Ajaib, I.~Gogoladze and Q.~Shafi,
  Phys.\ Rev.\ D {\bf 86}, 095028 (2012).


\bibitem{Kang:2012sy}
  Z.~Kang, J.~Li and T.~Li,
  JHEP {\bf 1211}, 024 (2012).


\bibitem{Hall}
L. J. Hall, D. Pinner and J. T. Ruderman, JHEP 1204, 131 (2012),
1112.2703; 
  B.~Kyae and J.~-C.~Park,
  Phys.\ Rev.\ D {\bf 87}, 075021 (2013);  
 T. Gherghetta, B. von Harling, A. D. Medina and M. A.
Schmidt, JHEP 1302, 032 (2013).

\bibitem{PQ:mass}
D. J. Miller, R. Nevzorov and P. M. Zerwas, 
 Nucl. Phys. B 681, 3 (2004).


\bibitem{Kang:2013rj}
  Z.~Kang, J.~Li, T.~Li, D.~Liu and J.~Shu,
  arXiv:1301.0453.

\bibitem{Christensen:2013dra}
  N.~D.~Christensen, T.~Han, Z.~Liu and S.~Su,
  arXiv:1303.2113.

\bibitem{Choi:2012he}
  K.~Choi, S.~H.~Im, K.~S.~Jeong and M.~Yamaguchi,
  arXiv:1211.0875; 
  C.~Cheung, S.~D.~McDermott and K.~M.~Zurek,
  JHEP {\bf 1304}, 074 (2013); 

\bibitem{Chang}
S. Chang, P.J. Fox and N. Weiner, 
JHEP 08 (2006) 068; 
 R. Dermisek and J.F. Gunion, 
Phys. Rev. D 77 (2008) 015013. 

\bibitem{Cao:2012fz}
 J.~-J.~Cao, Z.~-X.~Heng, J.~M.~Yang, Y.~-M.~Zhang and
J.~-Y.~Zhu,
  JHEP {\bf 1203}, 086 (2012); 
  K.~S.~Jeong, Y.~Shoji and M.~Yamaguchi,
  JHEP {\bf 1209}, 007 (2012); 
  K.~Agashe, Y.~Cui and R.~Franceschini,
  arXiv:1209.2115; 
  T.~Cheng and T.~Li,
  arXiv:1305.3214; 
  S.~F.~King, M.~M¨¹hlleitner, R.~Nevzorov and K.~Walz,
  Nucl.\ Phys.\ B {\bf 870}, 323 (2013); 



\bibitem{Cerdeno:2013cz}
  D.~G.~Cerdeno, P.~Ghosh and C.~B.~Park,
  JHEP {\bf 1306}, 031 (2013); 
  R.~Barbieri, D.~Buttazzo, K.~Kannike, F.~Sala and A.~Tesi,
  arXiv:1307.4937; 
  U.~Ellwanger,
  arXiv:1306.5541; 
  D.~G.~Cerdeno, P.~Ghosh, C.~B.~Park and M.~Peiro,
  arXiv:1307.7601; 
  J.~Hasenkamp and M.~W.~Winkler,
  arXiv:1308.2678.




\bibitem{Ellwanger}
U. Ellwanger, 
arXiv:1112.3548; 
  R.~Benbrik, M.~Gomez Bock, S.~Heinemeyer, O.~Stal, G.~Weiglein and L.~Zeune,
  Eur.\ Phys.\ J.\ C {\bf 72}, 2171 (2012); 
 C.~Beskidt, W.~de Boer and D.~I.~Kazakov,
  arXiv:1308.1333.


\bibitem{Schael:2006cr}
  S.~Schael {\it et al.}  [ALEPH and DELPHI and L3 and OPAL and LEP Working Group for Higgs Boson Searches Collaborations],
  Eur.\ Phys.\ J.\ C {\bf 47}, 547 (2006).

\bibitem{Djouadi:1997yw}
  A.~Djouadi, J.~Kalinowski and M.~Spira,
  Comput.\ Phys.\ Commun.\  {\bf 108}, 56 (1998)
.

 \bibitem{Belyaev:2012qa}
  A.~Belyaev, N.~D.~Christensen and A.~Pukhov,
  Comput.\ Phys.\ Commun.\  {\bf 184}, 1729 (2013).

\bibitem{NMSSMTools}
U. Ellwanger and C. Hugonie, Comput. Phys. Commun. 175 (2006) 290;
U. Ellwanger, J. F. Gunion, and C. Hugonie, JHEP 02 (2005) 066.



\bibitem{Aparicio:2012vk}
  L.~Aparicio, P.~G.~Camara, D.~G.~Cerdeno, L.~E.~Ibanez and I.~Valenzuela,
  JHEP {\bf 1302}, 084 (2013).

\bibitem{st1}
  ATLAS Collaboration, ATLAS-CONF-2013-053.
\bibitem{st2}
  ATLAS Collaboration, ATLAS-CONF-2013-037;
  ATLAS Collaboration, ATLAS-CONF-2013-048;
  CMS Collaboration, CMS-PAS-SUS-13-011.


\bibitem{Yan}
  X.~-J.~Bi, Q.~-S.~Yan and P.~-F.~Yin,
  arXiv:1209.2703;
  J.~Cao, C.~Han, L.~Wu, J.~M.~Yang and Y.~Zhang,
  JHEP {\bf 1211}, 039 (2012);
T. Cheng, J. Li, T. Li and Q. -S. Yan, arXiv:1304.318.



\bibitem{Badziak:2013bda}
  M.~Badziak, M.~Olechowski and S.~Pokorski,
  arXiv:1304.5437.

\bibitem{Arhrib:2004tj}
  A.~Arhrib and R.~Benbrik,
  Phys.\ Rev.\ D {\bf 71}, 095001 (2005). 

\bibitem{Chatrchyan:2012paa}
  S.~Chatrchyan {\it et al.}  [CMS Collaboration],
  JHEP {\bf 1303}, 037 (2013).

\bibitem{Harigaya:2012ir}
  K.~Harigaya, S.~Matsumoto, M.~M.~Nojiri and K.~Tobioka,
  Phys.\ Rev.\ D {\bf 86}, 015005 (2012).


\bibitem{Berenstein:2012fc}
  D.~Berenstein, T.~Liu, E.~Perkins and ,
  arXiv:1211.4288 [hep-ph].

\bibitem{Ghosh:2013qga}
  D.~Ghosh,
  arXiv:1308.0320 [hep-ph].

\bibitem{2b:MET}
ATLAS Collaboration, ATLAS-CONF-2012-165; S. Chatrchyan et al. [CMS
Collaboration], arXiv:1303.2985.

\bibitem{Chakraborty:2013moa}
  A.~Chakraborty, D.~K.~Ghosh, D.~Ghosh and D.~Sengupta,
  arXiv:1303.5776.

\bibitem{Plehn:2010st}
  T.~Plehn, M.~Spannowsky, M.~Takeuchi and D.~Zerwas,
  JHEP {\bf 1010}, 078 (2010).

 \bibitem{Alwall:2011uj}
  J.~Alwall, M.~Herquet, F.~Maltoni, O.~Mattelaer and T.~Stelzer,
  JHEP {\bf 1106}, 128 (2011).

 \bibitem{Sjostrand:2006za}
T.~Sjostrand, S.~Mrenna and P.~Z.~Skands,
JHEP {\bf 0605}, 026 (2006).

\bibitem{deFavereau:2013fsa}
  J.~de Favereau, C.~Delaere, P.~Demin, A.~Giammanco, V.~Lema"tre, A.~Mertens and M.~Selvaggi,
  arXiv:1307.6346 [hep-ex].

\bibitem{atlas2ssl}
  ATLAS Collaboration, ATLAS-CONF-2013-007.

\bibitem{Beenakker:1996ed}
  W.~Beenakker, R.~Hopker and M.~Spira,
  hep-ph/9611232.

\bibitem{CNMSSM}
  J.~F.~Gunion, Y.~Jiang and S.~Kraml,
  Phys.\ Lett.\ B {\bf 710}, 454 (2012);  
  U.~Ellwanger and C.~Hugonie,
  Adv.\ High Energy Phys.\  {\bf 2012}, 625389 (2012); 
  K.~Kowalska, S.~Munir, L.~Roszkowski, E.~M.~Sessolo, S.~Trojanowski and Y.~-L.~S.~Tsai,
  arXiv:1211.1693; 
  D.~Das, U.~Ellwanger and A.~M.~Teixeira,
  JHEP {\bf 1304}, 117 (2013).

\bibitem{Liu:2012qu}
  C.~Han, X.~Ji, L.~Wu, P.~Wu and J.~M.~Yang,
  arXiv:1307.3790.

\bibitem{Cahill-Rowley:2013vfa}
  M.~Cahill-Rowley, J.~Hewett, A.~Ismail and T.~Rizzo,
  arXiv:1308.0297.

\bibitem{Kribs:2013lua}
  G.~D.~Kribs, A.~Martin and A.~Menon,
  arXiv:1305.1313.

\bibitem{Batell:2012mj}
  B.~Batell, D.~McKeen and M.~Pospelov,
  JHEP {\bf 1210}, 104 (2012); 
  D.~Bertolini and M.~McCullough,
  JHEP {\bf 1212}, 118 (2012). 


\bibitem{Chpoi:2013wga}
  S.~Chpoi, S.~Jung and P.~Ko,
  arXiv:1307.3948.

\bibitem{Lopez-Val:2013yba}
  D.~Lopez-Val, T.~Plehn and M.~Rauch,
  arXiv:1308.1979.


\bibitem{shifman}
M. A. Shifman, A. I. Vainshtein, M. B. Voloshin and V. I. Zakharov,
Sov. J. Nucl. Phys. 30, 711 (1979) [Yad. Fiz. 30, 1368 (1979)].








\end{thebibliography}
\end{document}